\pdfoutput=1
\newcommand*{\ATLASLATEXPATH}{}
\documentclass[USenglish,texlive=2016,cernpreprint]{\ATLASLATEXPATH atlasdoc}

\usepackage{\ATLASLATEXPATH atlaspackage}
\usepackage{\ATLASLATEXPATH atlasbiblatex}
\usepackage{placeins}
\usepackage[hion]{\ATLASLATEXPATH atlasphysics}
\usepackage{HION-2015-06-PAPER-defs}
\usepackage{multirow}
\usepackage{siunitx}

\AtlasTitle{Measurement of the suppression and azimuthal anisotropy of muons from heavy-flavor 
decays in Pb+Pb collisions at $\sqrt{s_{\mathrm{NN}}} = 2.76$~\TeV\ with the ATLAS detector} 

\author{The ATLAS Collaboration}
\date{\today}

\PreprintIdNumber{CERN-EP-2018-072}
\AtlasJournalRef{Phys.\ Rev.\ C 98 (2018) 044905}
\AtlasDOI{10.1103/PhysRevC.98.044905}

\AtlasAbstract{%
ATLAS measurements of the production of muons from heavy-flavor decays 
  in $\sqrt{s_{\mathrm{NN}}} = 2.76$~\TeV\ Pb+Pb collisions and 
  $\sqrt{s} = 2.76$~\TeV\ $pp$ collisions at the LHC are presented.
Integrated luminosities of 0.14~$\mathrm{nb}^{-1}$ and
  570~$\mathrm{nb}^{-1}$ are used for the Pb+Pb and $pp$ measurements, respectively,
which are performed over the muon transverse momentum range
  $4 < p_{\mathrm{T}} < 14$~\GeV\  and for five Pb+Pb centrality intervals. 
Backgrounds arising from in-flight pion and kaon decays, hadronic 
  showers, and mis-reconstructed muons are statistically removed 
  using a template-fitting procedure. 
The heavy-flavor muon differential cross-sections and per-event yields 
  are measured in $pp$ and Pb+Pb collisions, respectively. 
The nuclear modification factor $R_{\mathrm{AA}}$ obtained from
  these is observed to be 
  independent of $p_{\mathrm{T}}$, within uncertainties, and to be less 
  than unity, which indicates suppressed production of heavy-flavor muons 
  in Pb+Pb collisions. For the 10\% most central Pb+Pb events, the measured $R_{\mathrm{AA}}$ 
  is approximately 0.35. 
The azimuthal modulation of the heavy-flavor muon yields is also measured
  and the associated Fourier coefficients $v_n$ for $n {=} 2$, 3 and 4 are given as a 
  function of $p_{\mathrm{T}}$ and centrality. 
They vary slowly with $p_{\mathrm{T}}$ and show a systematic variation 
  with centrality which is characteristic of other anisotropy measurements,
  such as that observed for inclusive hadrons.
The measured $R_{\mathrm{AA}}$ and $v_n$ values are also compared with 
  theoretical calculations.
}

\hypersetup{pdftitle={ATLAS draft},pdfauthor={The ATLAS Collaboration}}

\begin{document}
\maketitle

\section{Introduction}
\label{sec:intro}
Heavy quarks, especially bottom quarks, provide an important probe of
  the properties of the quark--gluon plasma created in high-energy
  nuclear (\NucNuc) collisions~\cite{vanHees:2004gq,vanHees:2005wb,Herzog:2006gh,vanHees:2007me,Horowitz:2007su,Uphoff:2011ad,He:2011qa,Cao:2013ita,}. 
The masses of the charm and bottom quarks are much larger than the temperatures 
  of 200--500~\MeV\ attained in the plasma (Ref.~\cite{Abreu:2007kv} 
  and references therein). 
As a result, the heavy quarks are mostly produced
  early in the collision at rates that are, in principle, calculable
  using perturbative QCD, and their subsequent interactions with the
  plasma give experimentally observable signatures. 
At transverse momenta (\pT) much greater than the mass of the bottom quark, 
  heavy quarks are expected to lose energy similarly to light quarks but with 
  mass-dependent modifications to the pattern of collisional and radiative 
  energy loss~\cite{Dokshitzer:2001zm,Djordjevic:2003zk,Armesto:2003jh,Djordjevic:2005db,Herzog:2006gh,Gossiaux:2010yx,Gossiaux:2009hr,}. 
At lower transverse momenta, $\pT \lesssim m_b$, the quarks are expected
  to diffuse in the plasma~\cite{CasalderreySolana:2006rq,vanHees:2007me,He:2011qa,}, 
  losing energy and partially thermalizing~\cite{Moore:2004tg,vanHees:2004gq}. 
As a result of their interactions with the collectively expanding medium, 
  the heavy quarks may acquire an azimuthal anisotropy.
Previous measurements of heavy-flavor production in \NucNuc\ collisions at RHIC 
  and the LHC, using semi-leptonic decays~\cite{Adare:2006nq,Adare:2010de,Abelev:2012qh,Adam:2015pga,} 
  and direct reconstruction of heavy-flavor mesons~\cite{Adamczyk:2014uip,Adamczyk:2017xur,ALICE:2012ab, Abelev:2013lca,Abelev:2014ipa,}, 
  have shown both substantial suppression in the yield of heavy quarks due to 
  energy loss and significant azimuthal anisotropy. 
Measurements of the heavy-quark yield and azimuthal anisotropy in \PbPb\ 
  collisions at the LHC can provide valuable constraints on plasma transport 
  parameters, such as the heavy-quark diffusion coefficient, and potentially 
  distinguish between weak- and strong-coupling models for heavy-quark 
  interactions in the plasma~\cite{Horowitz:2007su,CaronHuot:2008uh,Gubser:2009sn,HoyosBadajoz:2009pv,Noronha:2010zc,Chernicoff:2012iq,}.

The yield of particles produced in hard-scattering processes in \NucNuc\ 
  collisions is often characterized using the nuclear modification factor
\begin{equation}
\RAA = \frac{1}{\ATAA} \frac{ \left.\frac{1}{\Nevt}
\frac{d^2 N}{d\pt d\eta} \right
|_{\mathrm{cent}}} {\frac{d^2
\sigma^{pp}}{d\pt d\eta} }\,,
\label{eqn:raa_def}
\end{equation}
where $\eta$ is the pseudorapidity, the numerator is the differential per-event 
  yield in \NucNuc\ collisions for a given centrality interval, the denominator 
  is the \pp\ differential cross-section for producing the given particles, and
  \ATAA\ represents the nuclear overlap function averaged over the
  centrality interval~\cite{Glauber:1970jm}. 
In the absence of significant modification to the nuclear parton distributions 
  and of final-state interactions of the outgoing partons, 
  \RAA\ should be unity. 
Measurements of the production of vector bosons \cite{HION-2012-07,Chatrchyan:2012nt,Chatrchyan:2012vq,HION-2013-05,HION-2012-02} in \PbPb\ collisions at the LHC
  have verified this expectation. 
In contrast, measurements of \RAA\ for jets \cite{HION-2013-06,Adam:2015ewa} 
  and single hadrons~\cite{CMS:2012aa,Abelev:2012hxa,HION-2011-03} have shown a  
  centrality-dependent suppression that is understood to result from the
  energy loss of the parent quarks and gluons 
  (Refs.~\cite{Gyulassy:2003mc,Majumder:2010qh,CasalderreySolana:2007pr,} and
  references therein). 
Measurements of $D$-meson production in \PbPb\ collisions at the 
  LHC~\cite{ALICE:2012ab} have shown a centrality- and \pT-dependent 
  suppression similar to that observed for single hadrons. 
A measurement of $b$-hadron production, via their inclusive decays to 
  $J/\psi$ mesons, has also shown significant suppression~\cite{CMS-HIN-10-006}. 
Separate measurements of the production of forward heavy-flavor electrons~\cite{Adam:2016khe} 
  and muons~\cite{Abelev:2012qh} that are predominantly produced in 
  semi-leptonic $B$- and $D$-meson decays give \RAA\ values that are 
  significantly larger than those observed for inclusive hadrons. 
However, the $b\rightarrow$$J/\psi X$ and forward muon measurements are 
  statistically limited and insufficient to test theoretical calculations.

The azimuthal anisotropy of particles produced in an \NucNuc\ collision is
  often characterized by harmonic coefficients $v_n$ in a Fourier expansion of
  the particle yield as a function of azimuthal angle $\phi$~\cite{Poskanzer:1998yz} 
\begin{equation}
\displaystyle \frac{dN}{d\phi} = \left\langle \frac{dN}{d\phi} \right\rangle
\left(1 +  2\sum_{n\geq 1}  v_n \cos{\left(n\left[\phi - \Phi_n\right]\right)}\right),
\label{eq:flow}
\end{equation}
where $\Phi_n$ represents the event-plane angle for the $n$-th harmonic.
In non-central collisions, the azimuthal anisotropy is usually dominated by 
  the $n = 2$ term due to the almond-like shape of the collision geometry in the 
  transverse plane resulting from the non-zero impact parameter. 
Measurements of inclusive~\cite{Aamodt:2010pa,HION-2011-05,Aamodt:2011by,HION-2011-01,Chatrchyan:2013nka,} 
  and identified hadron~\cite{Abelev:2014pua,Abelev:2012di,} \vn\ values in
  \NucNuc\ collisions at the LHC and at RHIC show the presence of significant 
  azimuthal anisotropies, which are well reproduced by 
  hydrodynamic calculations.
These results provide the basis for the interpretation
  that the medium created in heavy-ion collisions is strongly coupled. 
The elliptic flow of heavy-flavor hadrons depends both on the coupling of the 
  heavy quark with the medium and on the transfer of the collective motion of 
  the medium to the heavy-flavor hadron in the hadronization process~\cite{Greco:2003vf}. 
The measurements of $D$-meson elliptic flow at mid-rapidity at the 
  LHC~\cite{Abelev:2013lca,Abelev:2014ipa} give \vtwo\ values that are similar 
  to those measured for light hadrons, while the forward-rapidity heavy-flavor 
  \vtwo\ values measured using semi-leptonic decays to muons are significantly 
  smaller. 
However, those measurements are statistically limited and, thus, do not provide 
  stringent constraints on theoretical calculations of the heavy-flavor 
  elliptic flow.

This \papertype\ presents ATLAS measurements of 
  muons from heavy-flavor semi-leptonic decays (heavy-flavor muons, hereafter)
  in \pp\ collisions at $\sqs = 2.76$~\TeV\ and \PbPb\ collisions at 
  $\sqn = 2.76$~\TeV. 
The \PbPb\ data were recorded during 2011, and the \pp\ data were recorded 
  during 2013.
The measurements are performed using data sets with integrated luminosities of
  570~\inb\ and 0.14~\inb\ for \pp\ and \PbPb\ collisions, respectively.
They are performed for several intervals of collision
  centrality, characterized using the total transverse energy measured
  in the forward calorimeters, and for different muon \pT\ intervals spanning
  the range 4--14~\GeV. 
Heavy-flavor muons are statistically separated from background muons resulting 
  from pion and kaon decays and from hadronic interactions using a 
  ``momentum-imbalance'' variable (Section~\ref{sec:prompt}) that compares the 
  momenta of the muons measured in the inner detector and muon spectrometer. 
Over the \pT\ range of the measurement, the residual irreducible 
  contamination by non-heavy-flavor muons, including contributions from \Jpsi\ 
  decays~\cite{STDM-2011-07,BPHY-2012-02},
  is less than 1\% and is neglected in the following.
The heavy-flavor muon differential per-event yields in \PbPb\ collisions 
  and differential cross-sections in \pp\ collisions measured over the 
  pseudorapidity interval $|\eta| < 1$ are used to calculate the heavy-flavor 
  muon \RAA\ as a function of \pT\ in different \PbPb\ centrality intervals. 
In addition, heavy-flavor muon \vn\ values are measured for $n$ = 2--4 
  as a function of \pT\ and collision centrality over $|\eta|<2$  using both 
  the event-plane and scalar-product~\cite{Luzum:2012da} methods. 
The scalar-product method has become the {\em de facto} standard procedure for
  \vn\ measurements using event-plane reconstruction. 
However, the method introduces additional complexity to the background 
  subtraction procedure (see Section~\ref{sec:vtwo_method}), so results 
  obtained using both methods are provided.
The results presented in this paper provide significantly improved
  statistical precision over previous measurements
  of the suppression and the anisotropic flow of semi-leptonically
  decaying heavy-flavor hadrons in \PbPb\ collisions at the LHC.

This \papertype\ is structured as follows.
  Section~\ref{sec:detector} describes the components of the ATLAS
  detector and trigger system used in the measurement,
  Section~\ref{sec:analysis} describes the data analysis,
  Section~\ref{sec:systematics} discusses the systematic uncertainties, and 
  the results are discussed in Section~\ref{sec:results}.
  Section~\ref{sec:conclusion} provides a summary and outlook.

\section{ATLAS detector}
\label{sec:detector}
The measurements presented in this \papertype\ use the
  ATLAS muon spectrometer (MS), inner detector (ID), calorimeter, 
  trigger and data acquisition systems. 
A detailed description of these detectors and their performance in 
  $pp$ collisions is given in Ref.~\cite{PERF-2007-01}. 
Muons are reconstructed by combining independent measurements of the muon 
  trajectories from the ID and the MS. 
The ID measures charged particles within the pseudorapidity interval\footnote{
  ATLAS uses a right-handed coordinate system with its origin at the nominal 
  interaction point (IP) in the center of the detector and the $z$-axis along 
  the beam pipe.   
  The $x$-axis points from the IP to the center of the LHC ring, and the $y$-axis points upward.  
  Cylindrical coordinates ($r$,$\phi$) are used in the transverse plane, $\phi$
  being the azimuthal angle around the $z$-axis.  
  The pseudorapidity is defined in terms of the polar angle $\theta$ as $\eta=-\ln\tan(\theta/2)$.
  } 
  $|\eta|<2.5$ using silicon pixel detectors, silicon microstrip detectors (SCT), and a
  straw-tube tracker, all immersed in a 2~T axial magnetic field. 
A charged particle typically traverses three layers of silicon pixel
  detectors, four layers of double-sided microstrip sensors, and 36 straws. 
The ID is surrounded by electromagnetic and hadronic calorimeters that absorb 
  efficiently the copious charged and neutral hadrons produced in Pb+Pb collisions. 
A muon typically loses 3 to 5~\GeV\ of energy, depending on the muon pseudorapidity,
  while crossing the calorimeters. 
The MS surrounds the calorimeters and provides tracking for muons within $|\eta|<2.7$ in the magnetic
  field produced by three air-core toroid magnet systems. 
Muon momenta are measured in the MS using three stations of precision drift chambers. 
Fast tracking detectors are used to trigger on muons in the MS.

Two forward calorimeters (FCal) are placed symmetrically
with respect to $z = 0$ and cover $3.2 < |\eta| < 4.9$. They are
composed of tungsten and copper absorbers with liquid argon as the
active medium; each calorimeter has a total thickness of about 10
interaction lengths.

Minimum-bias \PbPb\ collisions are identified using the zero-degree
calorimeters (ZDCs) and the minimum-bias trigger scintillator (MBTS)
counters~\cite{PERF-2007-01}. The ZDCs are located symmetrically at $z
= \pm 140$~m and cover $|\eta| > 8.3$. They are used only in
\PbPb\ collisions where they
primarily measure ``spectator'' neutrons, which originate from the
incident nuclei and do not scatter hadronically during the collision. 
The MBTS system detects charged particles over $2.1 < |\eta| < 3.9$ using two hodoscopes of 16 
  counters each, placed at $z = \pm 3.6$~m.  
The MBTS counters provide measurements of both the pulse heights and arrival times of ionization
  energy depositions in each hodoscope. 

The ATLAS trigger system~\cite{PERF-2011-02} consists of a first-level (L1)  trigger 
implemented using a combination of dedicated electronics with programmable logic, 
and a software-based high-level trigger (HLT). 
Data used for this analysis were selected using a combination of minimum-bias triggers, 
  which provided a uniform sampling of the \PbPb\ inelastic
  cross-section, and triggers that selected rare physics signatures such
  as muons. 
The measurements presented here are primarily obtained from
muon triggers. 
Events from the minimum-bias triggers are used only for cross-checks.

The muon triggers are formed using a combination of a L1 trigger
  and an HLT muon trigger whose configuration differed between
  \PbPb\ and \pp\ operation. 
For the \PbPb\ data, the L1 trigger selected events having a total transverse 
  energy of more than 50~\GeV, and the HLT trigger selected events containing 
  a track in the MS whose \pT, when corrected for the average muon energy loss 
  in the calorimeter, is greater than 4~\GeV. 
In \pp\ data, the muon trigger required a standalone muon track in the MS at L1, and a 
  muon track reconstructed using both the ID and MS with $\pT > 4$~\GeV\ at the HLT.  
The muon trigger was unprescaled throughout the \PbPb\ run and sampled essentially all 
  of the delivered luminosity.
In the \pp\ run, the trigger was prescaled such that it sampled $\sim$14\% (570~\inb) 
  of the $4~{\mathrm{pb}^{-1}}$ delivered luminosity.

\section{Data analysis}
\label{sec:analysis}
\subsection{Event selection}
Charged-particle tracks and vertices are reconstructed from hits in the 
  ID using a track reconstruction algorithm~\cite{STDM-2010-06} whose 
  configuration changed between the \pp\ and \PbPb\ measurements to 
  account for the high hit density in heavy-ion collisions~\cite{HION-2011-05}.  
To remove non-collision backgrounds, \PbPb\ events are required to have a 
  reconstructed primary vertex, at least one hit in each MBTS counter, 
  and a time difference between the two MBTS time measurements of less than 5~ns; 
  \pp\ events are required to have at least one reconstructed primary vertex.

The centrality of \PbPb\ collisions is characterized by \ETfcal, 
  the total transverse energy measured in the FCal~\cite{HION-2011-05}. 
For the results presented in this \papertype, the minimum-bias \ETfcal\ 
  distribution is divided into centrality intervals according to the 
  following percentiles of the \ETfcal\ distribution ordered from
  the most central to the most peripheral collisions: 0--10\%, 10--20\%,
  20--30\%, 30--40\%, and 40--60\%. 
A Glauber Monte-Carlo analysis~\cite{Miller:2007ri} 
  is used to estimate \ATAA\ for each of the centrality intervals~\cite{HION-2013-06}. 
The results are provided in Table~\ref{tab:glauber}. 
\begin{table} [!htb]
\caption{
The \ATAA\ values and their systematic uncertainties~\cite{HION-2013-06} in
  each centrality interval used in this analysis.
For the 40--60\% centrality interval, the \ATAA\ values are obtained 
  by averaging the values for 40--50\% and 50--60\% centrality intervals 
  from Ref.~\cite{HION-2013-06}.}
\centering
\begin{tabular}{|l|r|}\hline
Centrality interval [\%] & \ATAA\ [$\mathrm{mb}^{-1}$] \\ \hline
  0--10   & 23.45 $\pm$ 0.37  \\ \hline             
 10--20   & 14.43 $\pm$ 0.30  \\ \hline             
 20--30   & 8.73 $\pm$ 0.26   \\ \hline             
 30--40   & 5.04 $\pm$ 0.22   \\ \hline             
 40--60   & 2.02 $\pm$ 0.15   \\ \hline      
\end{tabular}
\label{tab:glauber}
\end{table}

\subsection{Muon reconstruction}
\label{sec:muonrec}
Muons in this analysis are formed by combining tracks
  reconstructed in the MS~\cite{STDM-2011-07} 
  with the tracks measured in the ID. 
The associated ID tracks are required to satisfy criteria for the number 
  of hits in the SCT and pixel detectors which are the same for the \pp\ 
  and \PbPb\ data, but which are optimized for the \PbPb\ analysis~\cite{HION-2011-05}.  
In particular, for both data sets, ID tracks are required to have 
  transverse and longitudinal impact parameters relative to the
  reconstructed primary vertex of less than 5~mm and to have a momentum 
  $p > 3$~\GeV.
The requirements on the longitudinal and transverse impact parameters are 
  relaxed to 5~mm, compared to the 1~mm (or 1.5~mm) typically used in heavy-ion 
  analyses~\cite{HION-2011-05,HION-2011-01}, to allow selection of muons 
  from off-vertex heavy-flavor decays.
The ID tracks are also required to have at least one pixel hit, with the additional 
  requirement of a hit in the first pixel layer when one is expected,\footnote{A hit 
     is expected if the extrapolated track crosses an active region of a pixel module 
     that has not been disabled.}
  at least seven SCT hits, and at most one hit that is expected but not found
  in the pixel and SCT detectors taken together.
The transverse momentum measured in the MS (\pMSt) is required to be greater than 1.2~\GeV\ 
  for both  the \pp\ and \PbPb\ data. 
In the \PbPb\ analysis, this selection removes muons for which 
  the \PbPb\ trigger efficiency is less than 50\%.

The results presented here use muons having $4 < \pT < 14$~\GeV\ 
  and having $|\eta| < 1$ for the heavy-flavor-suppression analysis
  or $|\eta| < 2$ for the flow measurements. 
The lower limit of the \pT\ range is constrained by the \pT\ dependence 
  of the muon trigger and reconstruction efficiencies, while the upper 
  limit is determined by the number of events available in the \PbPb\ data. 
For the \RAA\ measurements, a muon $\eta$ interval of $|\eta| < 1$ is chosen, 
  as the muon trigger and reconstruction have optimal performance over this $\eta$ range.
The $\eta$ range is extended to $|\eta| < 2$ for the \vn\ measurements, 
  as they are not sensitive to the effects of trigger and tracking efficiency.
A total of 9.2~million  (1.8~million) muons are reconstructed within 
  these kinematic ranges from 8.7 million (1.8 million) events recorded 
  using the \PbPb\ (\pp) muon triggers.

The performance of the ATLAS detector and offline analysis in measuring 
  muons in \pp\ collisions is evaluated by a \GEANT4~\cite{Agostinelli:2002hh} 
  simulation of the ATLAS detector~\cite{SOFT-2010-01} using Monte Carlo (MC) 
  $\sqs =$~2.76~\TeV\ \pp\  events produced with the \PYTHIA\ event 
  generator~\cite{Sjostrand:2006za} (version 6.423 with parameters chosen
  according to the AUET2B set of tuned parameters ~\cite{mctunes}).
The reconstruction performance in \PbPb\ collisions is evaluated by ``overlaying''
  simulated \PYTHIA\ \pp\ events 
  on minimum-bias \PbPb\ events.
In this overlay procedure, the simulated hits are combined with the data from minimum-bias
  events to produce the final sample.
The minimum-bias \PbPb\ events used in the overlay procedure were recorded by ATLAS 
  during the same data-taking period as the data used in this analysis. 
For both the \pp\ and \PbPb\ measurements, the muon reconstruction efficiency 
  increases by about 30\% from $\pT = 4$~\GeV\ to $\pT = 6$~\GeV, above 
  which it is approximately constant at 0.80 and 0.77 for the 
  \pp\ and \PbPb\ data, respectively. 
The \PbPb\ muon reconstruction efficiency is independent of the centrality 
  within uncertainties. 

The \PbPb\ muon trigger efficiency is measured for fully reconstructed
  muons using the minimum-bias \PbPb\ data set. 
The efficiency is evaluated as the fraction of reconstructed muons for which the HLT
  finds a matching muon with $\pT > 4$~\GeV.  
It is observed to be independent of centrality, within statistical uncertainties,
  and increases from about 0.6 at $\pT = 4$~\GeV\ to about 0.8 at 6~\GeV, above
  which it is approximately constant. 
The \pp\ muon trigger efficiency is similarly evaluated using \pp\ 
  events selected by a set of minimum-bias triggers.
The efficiency increases from 0.40 for $\pT = 4$~\GeV\ to $0.75$ for
  $\pT = 12$~\GeV.

\subsection{Heavy-flavor-suppression measurement}
\label{sec:prompt}
The muons measured in the \pp\ and \PbPb\ data sets contain background
  from in-flight decays of pions and kaons, muons produced from the
  decays of particles produced in hadronic showers in the material of 
  the detector, and mis-associations of ID and MS tracks. 
Previous studies have shown that the signal and background contributions 
  to the reconstructed muon sample can be  discriminated 
  statistically~\cite{STDM-2011-07}. 
This analysis relies solely on the fractional momentum imbalance 
  $\Deltaploss/\pID$, which quantifies the difference between the ID
  and MS measurements of the muon momentum after accounting for the
  energy loss of the muon in the calorimeters. 
It is defined as
\begin{eqnarray*}
\dfrac{\Deltaploss}{\pID} = \dfrac{\pID - \pMS - \Deltapcalo(p, \eta, \phi)}{\pID},
\end{eqnarray*}
where \pID\ and \pMS\ represent the reconstructed muon momenta from the ID
  and MS, respectively, and $\Deltapcalo$ represents the momentum- and
  angle-dependent average momentum loss of muons in the calorimeter
  obtained from simulations. 
Muons resulting from background processes typically have \pMS\ values 
  smaller than would be expected for a muon produced directly in \pp\ 
  or \PbPb\ collisions or via the decays of heavy-flavor hadrons. 
This is because the background muons from pion/kaon decays or from 
  hadronic interactions in the calorimeter have, on average, 
  smaller \pt\ compared to the parent particle.
As a result, background muons are expected to have $\dpop > 0$. 
\begin{figure}[!t]
\centerline{
\includegraphics[width=0.9\columnwidth]{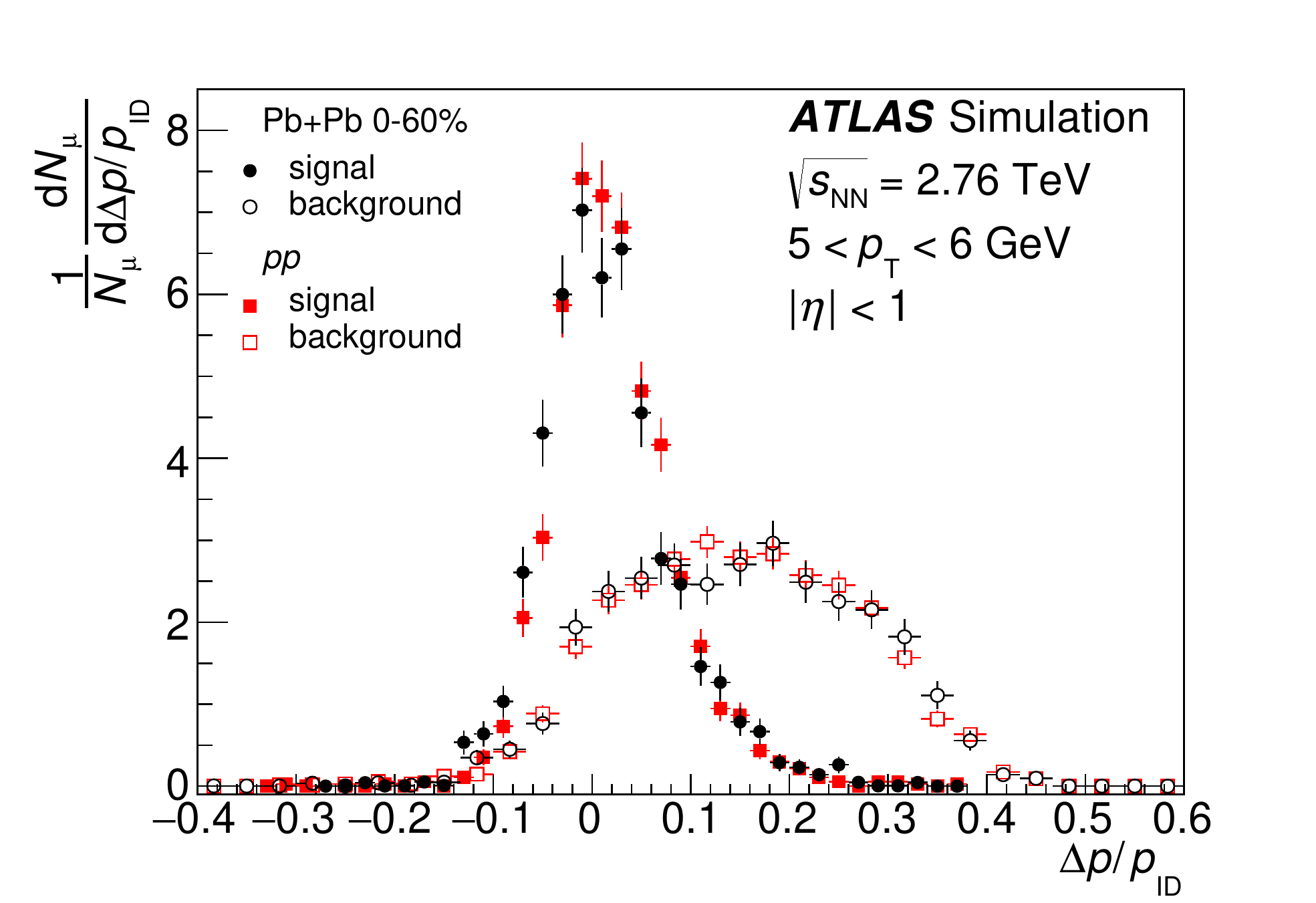}
}
\caption{
Signal and background template distributions in \pp\ 
  collisions (square points) and \PbPb\ collisions 
  (circular points) in the 0--60\% centrality interval for muons 
  having $5 < \pT < 6$~\GeV\ and $|\eta| < 1$. 
The signal and background distributions are separately normalized such
  that their integral is unity. 
For clarity, the background distribution is binned more coarsely.
} 
\label{fig:ploss}
\end{figure}

Distributions for \dpop\ are obtained from the simulated samples separately
  for signal muons and for background muons. 
The signal muons include muons directly produced 
  in electromagnetic decays of hadrons, 
  in decays of $\tau$-leptons,
  in decays of $W$ and $Z$ bosons,
  in decays of top quarks,
  and in semi-leptonic decays of heavy-flavor hadrons; this last 
  contribution dominates the signal sample, contributing about 99\% 
  of the muons over the \pt\ range measured in this analysis (Ref.~\cite{STDM-2011-07}
  and references therein).
The different contributions to the background -- pion decays in flight, 
  kaon decays in flight, muons produced  by secondary interactions of 
  prompt particles, and mis-associations -- are evaluated separately. 
Figure~\ref{fig:ploss} shows MC distributions of \dpop\ for signal and 
  background muons having $5 < \pT < 6$~\GeV\ for \PbPb\ collisions in 
  the centrality range 0--60\% and for \pp\ collisions. 
The \dpop\ distribution for signal muons is centered at zero while the 
  distribution for background muons is shifted to positive values. 
The signal distributions show only modest differences between \pp\ and
  \PbPb\ collisions. 
Similarly, when making separate templates for different \PbPb\ collision centralities, 
  a weak dependence of the signal templates on centrality is observed. 
The background \dpop\ distributions are much
  broader and are insensitive to the centrality-dependent 
  effects seen in the signal distributions.

A template-fitting procedure is used to estimate statistically the
  signal fraction for each kinematic and centrality selection 
  used in the analysis. 
The measured \dpop\ distribution is assumed to result from a combination 
  of signal and background distributions
\begin{eqnarray*}
\frac{1}{N_{\mathrm{\mu}}}\frac{d N_{\mu}}{d\dpop} =  \fpr
\frac{d P^{\mathrm{sig}}}{d\dpop} + \left(1 -
\fpr\right)\frac{d P^{\mathrm{bkg}}}{d\dpop},
\end{eqnarray*}
where $N_{\mathrm{\mu}}$ is the total number of muons in the sample,
  $d P^{\mathrm{sig}}/{d\dpop}$ and $d P^{\mathrm{bkg}}/d\dpop$ represent the signal and background
  \dpop\ probability distributions, respectively, and \fpr\ represents
  the signal fraction.

\begin{figure}[!t]
\centerline{
\includegraphics[width=0.99\columnwidth]{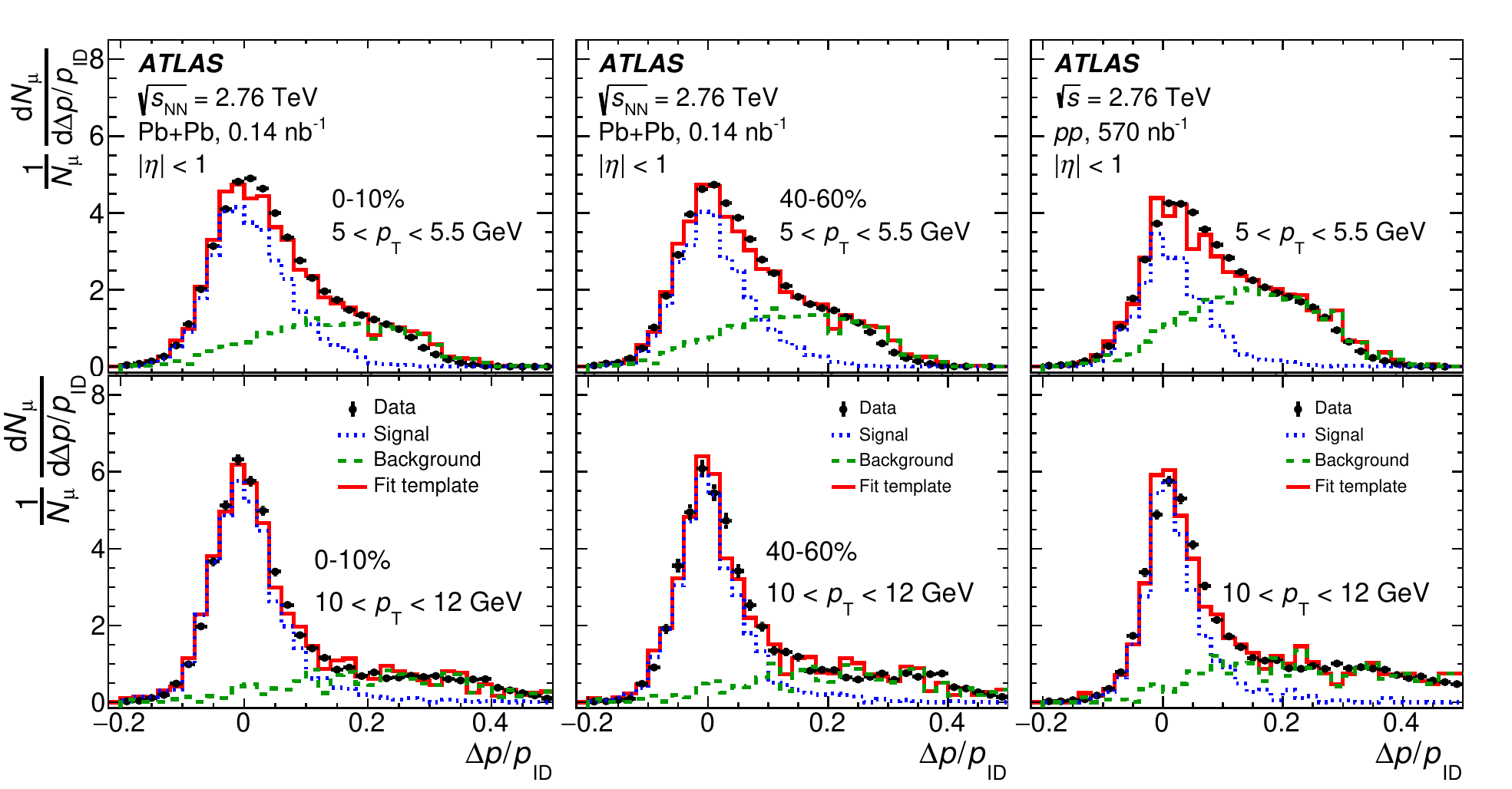}
}
\caption{
Examples of template fits to  \PbPb\ and \pp\ data.
The top panels show results for $5 < \pT < 5.5$~\GeV\ and the bottom panels
  show results for $10 < \pT < 12$~\GeV. 
The left, middle, and right panels show results for \PbPb\ 0--10\%, 
  \PbPb\ 40--60\%, and \pp, respectively.  
The black points represent the data. 
The dotted and dashed lines represent the signal and background template 
  distributions weighted by \fpr\ and $(1 - \fpr)$, respectively 
  (see text) and the continuous  lines represent the summed template distributions. 
} 
\label{fig:fits}
\end{figure}

For \PbPb\ data, centrality-dependent templates are used for the signal 
  while centrality-integrated templates are used for the background. 
The latter is motivated by the observed centrality independence of the 
  background templates and the limited size of the background sample. 
Template fits are performed using binned $\chi^2$ fits that account for 
  the statistical precision of the signal and background templates. 
The fits are performed using MINUIT~\cite{James:1975dr} with \fpr\ 
  as the free parameter.
The uncertainties from the fits are used as statistical uncertainties
  of the yields and propagated into the final results.
Example template fits are shown for two muon \pT\ intervals in 
  Figure~\ref{fig:fits} for \PbPb\ events in the 0--10\% and 40--60\% 
  centrality intervals and for \pp\ data. 
As shown in Figure~\ref{fig:fits}, the measured \dpop\ distributions are well described by 
a combination of the signal and background templates,
  and this holds for all studied kinematic and centrality intervals.

\begin{figure}[!t]
\centerline{
\includegraphics[width=0.99\columnwidth]{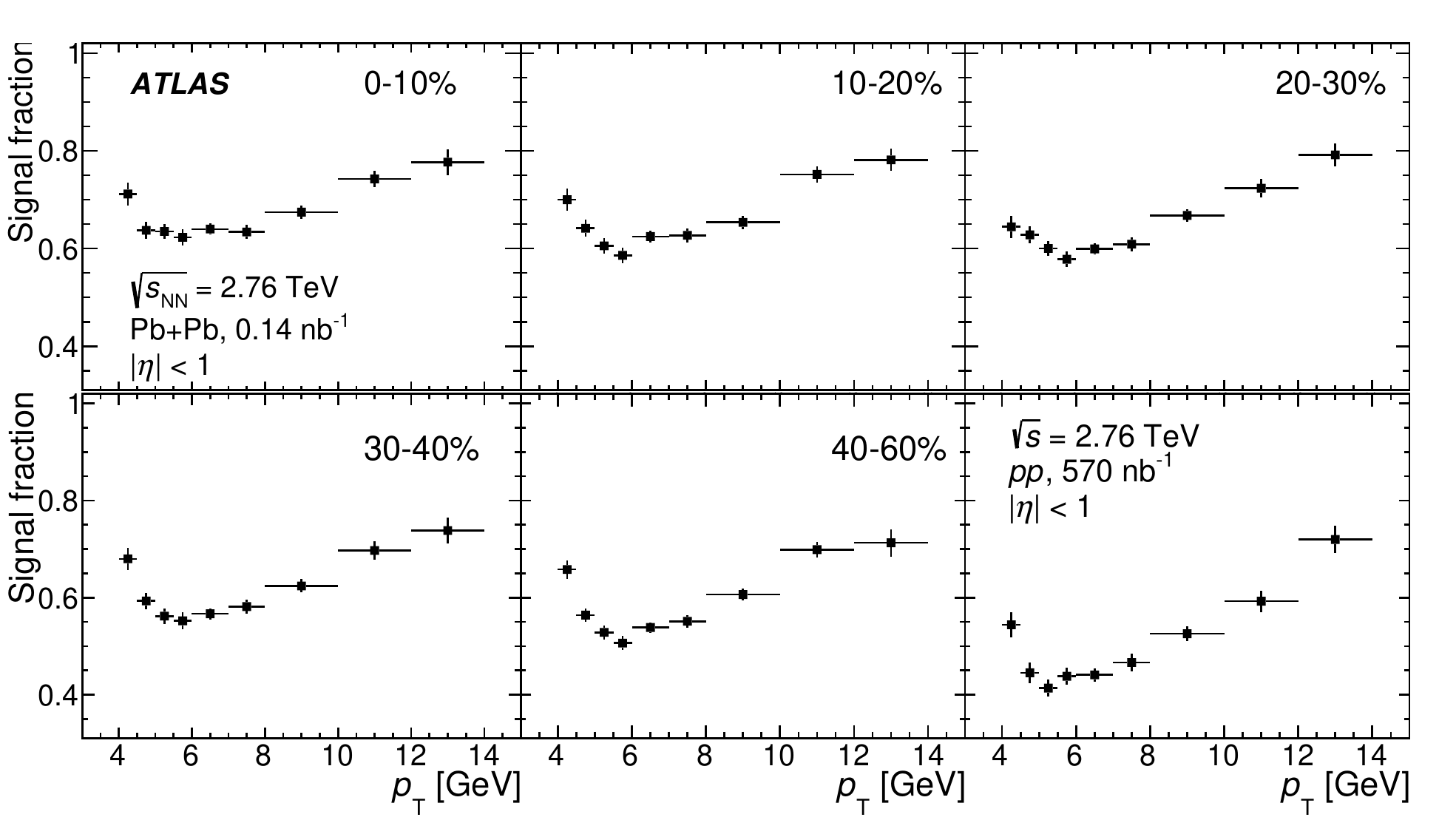}
}
\caption{
Signal fraction values obtained from template fits to the
  \PbPb\ and \pp\ data as a function of the muon \pT. 
Results are shown for different \PbPb\ centrality intervals and 
  for \pp\ collisions in the bottom right panel.
The error bars correspond to statistical uncertainties only. 
} 
\label{fig:fprRAA}
\end{figure}

The signal fractions \fpr\  obtained from the template fits using these intervals 
  are shown in Figure~\ref{fig:fprRAA} for the \PbPb\ and \pp\ data.
The signal fractions increase with \pt\ for $\pt > 5$~\GeV,  indicating that
  at higher \pt\ a larger fraction of the reconstructed muons are heavy-flavor (HF) muons.
The increase in \fpr\ at low \pT\ results from the trigger, which is less efficient 
  for background muons that have low \pMSt. 
Such an increase is not observed when repeating this analysis using 
  the minimum-bias \PbPb\ data set.
This increase in the \fpr\ due to the trigger does not affect the measurement, 
  as is demonstrated by studies of variations in the \pMSt\ criterion 
  in Section~\ref{sec:raa_systematics}.

With the \fpr\ obtained from the template fits, the \pp\ differential 
  cross-section for producing heavy-flavor muons is calculated according to
\begin{equation}
\frac{d^2 \sigma_{\mathrm{HF} \mu}}{d \pT d \eta} =
\frac{1}{\mathcal{L}} \frac{\Delta N_{\mu} \fpr}{\Delta \pT \Delta
    \eta} \frac{1}{\varepsilon_{\mathrm{trig}}
  \varepsilon_{\mathrm{rec}}} \, ,
\label{eq:ppsig} 
\end{equation}
where $\mathcal{L}$ is the integrated luminosity of the \pp\ measurement, 
  $\Delta\pt$ is the width of the given \pt\ interval,
$\Delta \eta = 2$ is the size of the pseudorapidity interval,
  $\Delta N_{\mu}$ represents the number of muons in the given \pT\ and $\eta$ intervals, 
  and $\varepsilon_{\mathrm{trig}}$ and $\varepsilon_{\mathrm{rec}}$ 
  represent the trigger and reconstruction efficiencies, respectively. 
The luminosity is calibrated using a set of beam-separation scans performed 
  in February 2013. 
It has a relative uncertainty of 3.1\% that was derived following a methodology
  similar to that detailed in Ref.~\cite{DAPR-2011-01}. 

The \PbPb\ differential per-event yields for producing 
  heavy-flavor muons are calculated according to
\begin{equation}
\left.\frac{1}{\Nevt} \frac{d^2 N_{\mathrm{HF} \mu}}{d \pT d \eta}\right|_{\mathrm{cent}} =
\frac{1}{N_{\mathrm{evt}}^{\mathrm{cent}}} \frac{\Delta N_{\mu}^{\mathrm{cent}} \fpr}{\Delta \pT \Delta
    \eta} \frac{1}{\varepsilon_{\mathrm{trig}}
  \varepsilon_{\mathrm{rec}}} \, ,
\label{eq:yieldPbPb} 
\end{equation}
where $N_{\mathrm{evt}}^{\mathrm{cent}}$ is the number of
  \PbPb\ collisions in a given centrality interval, 
  $\Delta N_{\mu}^{\mathrm{cent}}$ represents the number
  of total muons with $|\eta| < 1$ measured in the given \pT\ 
  and centrality interval, 
  \fpr\ represents the corresponding signal fraction obtained from
  the template fits, 
  and $\varepsilon_{\mathrm{trig}}$ and $\varepsilon_{\mathrm{rec}}$ 
  represent the trigger and reconstruction efficiencies, respectively.

\subsection{Azimuthal anisotropy measurement\label{sec:vtwo_method}}

The \vn measurements additionally require determination of the event-plane 
  (EP) angles $\Phi_n$ (Eq.~\eqref{eq:flow}). 
However, due to detector acceptance effects and finite particle multiplicity 
  in an event, the measured EP angles, denoted \psin, fluctuate 
  event-by-event around the true EP angles~\cite{Poskanzer:1998yz}. 
  The ``observed'' \vn, $\vn^{{\mathrm{obs}}}$, is obtained by measuring the distribution
  of the  particle directions relative to the \psin\ planes:
\begin{equation}
\frac{d N}{d\phi}=N_0\left[1+2\sum_{n\geq1}\vn^{\mathrm{obs}}\cos(n(\phi-\psin))\right].
\label{eq:v2_eqn}
\end{equation}
The $\vn^{\mathrm{obs}}$ are smaller in magnitude than the true \vn\ because 
  they are calculated around the \psin\ planes rather than the $\Phi_n$ planes.
To account for this, the $\vn^{\mathrm{obs}}$ are corrected by the 
EP resolution factor ${\mathrm{Res}}\{n\psin\}$, which accounts for the smearing 
  of \psin relative to $\Phi_n$~\cite{Poskanzer:1998yz}:
\begin{equation}
  \vn=\frac{\vn^{\mathrm{obs}}}{{\mathrm{Res}\{n\psin\}}},\ \ \ \  {\mathrm{Res}\{n\psin\}}=\langle{\cos(n(\psin-\Phi_{n}))}\rangle_{\mathrm{evts}},
\label{eq:reso_correction}
\end{equation}
where, the $\langle...\rangle_{\mathrm{evts}}$ indicates averaging over all 
  events in a given centrality class.
In this analysis, the \psin\ angle is determined 
  using the flow vector or ``$q$-vector'' method~\cite{Poskanzer:1998yz},
  in which the $q$-vector is calculated from the \ET\ deposited in the FCal according to:
\begin{eqnarray}
  q_{n,x}& = & \frac{\Sigma E_{{\mathrm{T}},i}\cos(n\phi_{i})-\langle\Sigma E_{{\mathrm{T}},i}\cos(n\phi_{i})\rangle_{{\mathrm{evts}}}}{\Sigma E_{{\mathrm{T}},i}},\label{eq:q_vec_defn}\\
  q_{n,y}& = & \frac{\Sigma E_{{\mathrm{T}},i}\sin(n\phi_{i})-\langle\Sigma E_{{\mathrm{T}},i}\sin(n\phi_{i})\rangle_{{\mathrm{evts}}}}{\Sigma E_{{\mathrm{T}},i}},\nonumber
\end{eqnarray}
where the sum is over all the calorimeter towers\footnote{
    Calorimeter towers are localized groups of calorimeter cells that have
    a $\delta\eta\times\delta\phi$ segmentation of $0.1\times0.1$.
  } 
  in the FCal, $E_{\mathrm{T},i}$ is the transverse energy deposited in 
  the $i^{\mathrm{th}}$ tower, and $\phi_{i}$ denotes the azimuthal angle
  of the position of the center of the tower.
The event-averaged terms $\langle\Sigma E_{\mathrm{T},i}\cos(n\phi_{i})\rangle_{{\mathrm{evts}}}$
  and $\langle\Sigma E_{\mathrm{T},i}\sin(n\phi_{i})\rangle_{{\mathrm{evts}}}$
  are subtracted in order to remove detector effects~\cite{Afanasiev:2009wq}.
From the $q_n$-vectors, the EP angles \psin, are determined as~\cite{HION-2012-03}
\begin{eqnarray*}
\tan(n\Psi_n)=\frac{q_{n,y}}{q_{n,x}}.
\end{eqnarray*}
The parameter ${\mathrm{Res}}\{n\Psi_{n}\}$ is determined by the two-subevents (2SE) 
  method~\cite{Poskanzer:1998yz}.
In the 2SE method, the signal from a detector used to measure the event plane is 
  divided into two ``subevents'' covering equal pseudorapidity ranges in
  opposite $\eta$ hemispheres, such that the two subevents nominally have the 
  same resolution.  
The FCal detectors located at positive and negative $\eta$,
  $\mathrm{FCal^P}$ and $\mathrm{FCal^N}$, provide such a division.  
The resolution of the $\mathrm{FCal^{P(N)}}$ is calculated from the correlation 
  between the two subevents
\begin{eqnarray*}
  \mathrm{Res}(n\Psi_n^{\mathrm{P(N)}})=\sqrt{\langle \cos n(\Psi_n^{\mathrm{P}} -\Psi_n^{\mathrm{N}}) \rangle},
\end{eqnarray*}
where $\Psi_n^{\mathrm{P(N)}}$ is the event-plane angle determined from the 
  positive (negative) side of the FCal.
From the subevent resolution the full FCal resolution can be determined 
  by the procedure described in Ref.~\cite{Poskanzer:1998yz}.
The ${\mathrm{Res}}\{n\Psi_{n}\}$ for the FCal and their associated systematic 
  uncertainties were determined in a previous 
  ATLAS analysis~\cite{HION-2011-01}.
Those values and uncertainties are directly used in this \papertype.
Depending on  the centrality class, the EP resolution factor for the FCal 
  varies between 0.7--0.9, 0.3--0.65, and 0.2--0.4 for \vtwo, \vthree\ 
  and \vfour, respectively. 
The uncertainties in the EP resolution factor are less than 3\%, 4\% and 6\% for \vtwo,
  \vthree\ and \vfour, respectively, for all the centrality classes used in this analysis.

The heavy-flavor muon $v_n^{\mathrm{obs}}$ values are measured by evaluating 
  the yields differentially relative to the \psin\ plane. 
For this, the template-fitting procedure is repeated in intervals of 
  $n|\phi-\psin|$ for each \pt\ and centrality interval. 
Utilizing the $n$-fold symmetry of the $\psin$ plane and the fact that 
  $\cos(n(\phi-\psin))$ is an even function, it is sufficient to bin only
  over the interval $(0,\pi)$ in $n|\phi-\psin|$.
Four intervals of $n|\phi-\psin|$ [(0,$\pi$/4), ($\pi$/4,$\pi$/2), ($\pi$/2,3$\pi$/4), 
  and (3$\pi$/4,$\pi$)] are used.
The same signal and background templates are used for the four
  $n|\phi-\psin|$ intervals in a given \pt\ and centrality interval. 
As a result, there is a significant correlation between the statistical
  uncertainties of the signal fractions measured in the
  four $\cos(n(\phi-\Psi_2))$ intervals. 
This correlation is accounted for in the statistical uncertainties 
  of the final \vn\ values.

\begin{figure}[!h]
\centerline{
\includegraphics[width=0.99\columnwidth]{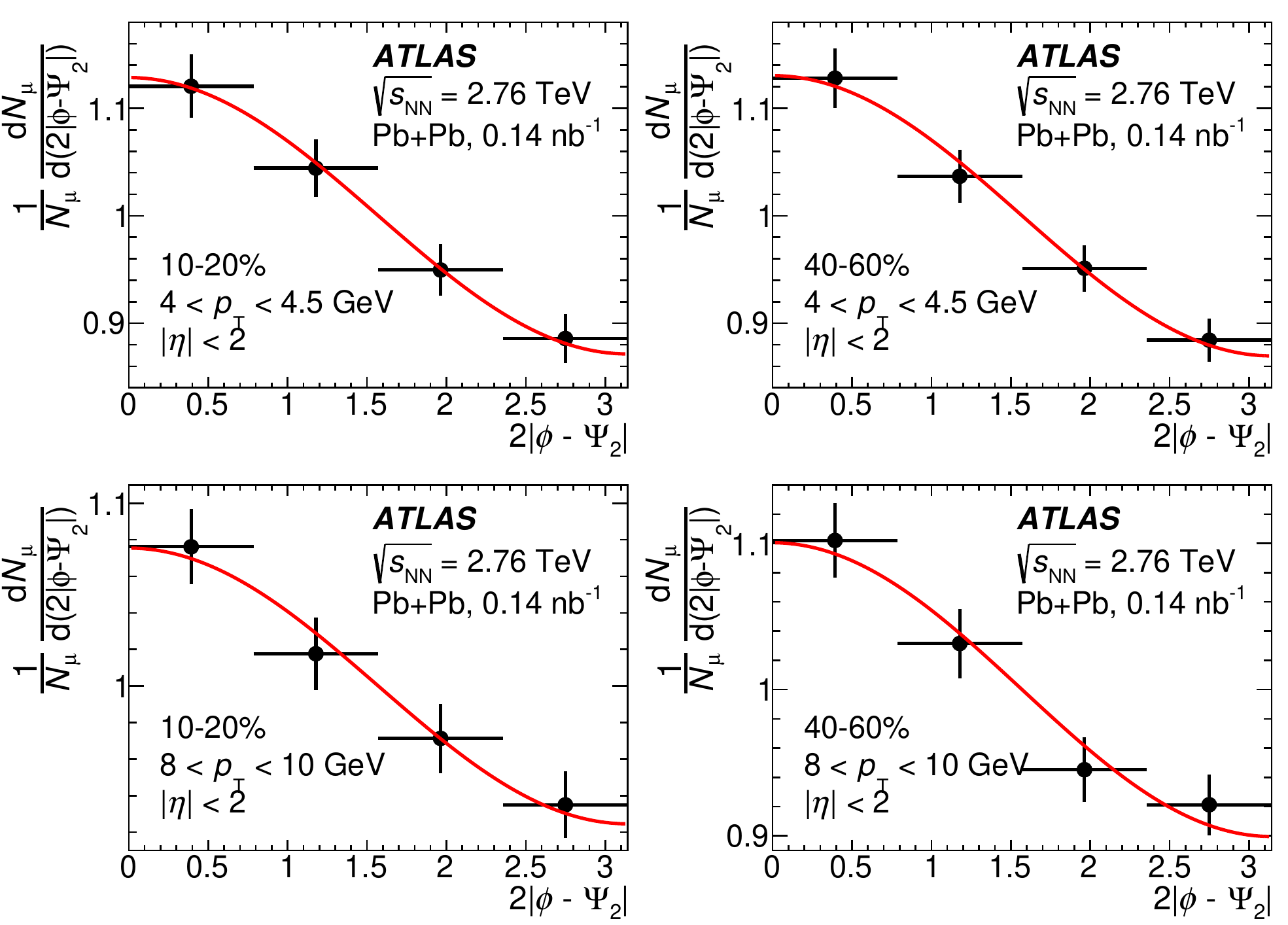}
}
\caption{
Examples of heavy-flavor muon yields, expressed in thousands of muons, as a
  function of $2|\phi-\psitwo|$ in intervals of $\pi/4$. 
The left  and right columns show results for the 10--20\% and 40--60\% centrality 
  intervals, respectively, and the top and bottom rows correspond to
  $4.0<\pt<4.5$~\GeV\ and $8<\pt<10$~\GeV, respectively. 
The error bars on the data points show statistical uncertainties from the fits.
There are significant bin-to-bin correlations between the statistical
  uncertainties due to the use of the same signal and background templates 
  in all $2|\phi-\psitwo|$ intervals.
The continuous lines indicate the results of fits of the data to Eq.~\eqref{eq:v2_eqn}.
} 
\label{fig:v2_fit}
\end{figure}

Figure~\ref{fig:v2_fit} shows examples of the differential yields of 
  heavy-flavor muons obtained from the template fits as a 
  function of $2|\phi-\psitwo|$ for two centrality and two \pT\ intervals.  
A clear dependence of the  yields on $2|\phi-\psitwo|$ can be observed, with a 
  larger yield in the ``in-plane'' direction ($2|\phi-\psitwo|\sim0$) 
  compared to the ``out-of-plane'' direction ($2|\phi-\psitwo|\sim\pi$), 
  implying a significant \vtwo\ signal.
The differential yields are fitted with a second-order Fourier function 
  of the form in Eq.~\eqref{eq:v2_eqn} to obtain the $v_2^{\mathrm{obs}}$ values. 
In the fits, the $\chi^2$ minimization takes into 
  account the correlations between the statistical uncertainties of the
  yields in the different $2|\phi-\psitwo|$ bins. 
These fits are indicated by the continuous lines in Figure~\ref{fig:v2_fit}. 
The $v_2^{\mathrm{obs}}$ values are then  corrected to account for the 
  EP resolution (Eq.~\eqref{eq:reso_correction}) for the final results 
  presented in Section~\ref{sec:results}. 

One drawback of the EP method is that there is an ambiguity 
  in the interpretation of the \vn\ values obtained from it
  (from here on the \vn\ values obtained from the event-plane method are denoted by \vnep).
In the limit of perfect EP resolution, ${\mathrm{Res}}\{n\Psi_{n}\}\rightarrow 1$, 
  $\vnep\rightarrow\langle\vn\rangle$, while in the limit of poor resolution, 
  ${\mathrm{Res}}\{n\Psi_{n}\}\rightarrow 0$, $\vnep\rightarrow\sqrt{\langle\vn^2\rangle}$
  where the $\langle...\rangle$ indicates an average over all events~\cite{Luzum:2012da}.
In general, the \vn\ values measured with the EP method lie somewhere between $\langle\vn\rangle$ 
and $\langle\sqrt{\vn^2}\rangle$, depending on the value of the resolution.
For this reason, the scalar-product (SP) method is considered to be a superior
  measurement technique, as it always measures the 
  r.m.s.\ \vn\ value, i.e. $\sqrt{\langle\vn^2\rangle}$~\cite{Luzum:2012da}.
The ideal SP method entails weighting the contribution of each measured signal muon by the magnitude 
  of the $q$-vector (Eq.~\eqref{eq:q_vec_defn}) measured in the FCal, giving
\begin{equation}
\vnsp=\frac{\langle q_n \cos(n(\phi-\psin)) \rangle_{\mathrm{evts}}}{\mathrm{Res}^{\mathrm{SP}}\{n\psin\}},
\label{vn_sp}
\end{equation}
where $\mathrm{Res}^{\mathrm{SP}}\{n\psin\}$ is the resolution for 
  the SP method, given by
\begin{eqnarray*}
\mathrm{Res^{SP}}(n\Psi_n)=\sqrt{\langle q_{n}^{\mathrm{P}}q_{n}^{\mathrm{N}}\cos n(\Psi_n^{\mathrm{P}} -\Psi_n^{\mathrm{N}}) \rangle},
\end{eqnarray*}
where $q_{n}^{\mathrm{P(N)}}$ is the magnitude of the $n^{\mathrm{th}}$-order $q$-vector measured 
  in the positive $z$ (negative $z$) side of the FCal. 
Previous ATLAS measurements for inclusive charged particles show that \vnep\  values differ 
  by less than 5\% from the r.m.s.\ \vn\ values for \vtwo, and 
  harmonics of order $n\geq3$ are consistent with the r.m.s.\ \vn\ within systematic 
  uncertainties~\cite{HION-2012-10}.
However, Eq.~\eqref{vn_sp} cannot be directly used in the present analysis,
  since a priori it is not known whether a reconstructed muon is a signal or background muon;
  the number of signal muons is statistically extracted from the 
  momentum imbalance distributions.
Instead, the implementation of the SP method follows quite closely the EP method.
The template fits are done in four  intervals of $n|\phi-\psin|$ with each muon 
  weighted with the measured $q_n$ in that event.
These fits give the $q_n$-weighted signal muon yields in each 
$n|\phi-\psin|$ interval. 
These weighted yields are then fitted with $n^{\mathrm{th}}$-order 
Fourier functions, similar to Figure~\ref{fig:v2_fit}, to obtain the observed SP \vn\ values,
  which are then corrected by $\mathrm{Res}^{\mathrm{SP}}\{n\psin\}$ to
  obtain the \vnsp, presented later in Section~\ref{sec:results}.

While the SP method has advantages over the EP method,
  only a modified version of the SP method can be used in the present analysis.
Thus, the results obtained from both the SP and EP methods are presented.

\subsection{Jet bias in the \vn\ measurement}
The heavy-flavor muons measured in this analysis often result from 
  heavy-flavor jets that have an associated back-to-back recoil jet.
If the recoil jet is in the FCal, it can bias the  orientation 
  of the \psin\ to be aligned with the azimuthal angle of the muon,
  yielding a larger measured \vn.
This ``jet bias effect'' needs to be estimated and corrected for 
  in the measurement.
The magnitude of this effect is estimated using the simulated-data overlay
  events described in Section~\ref{sec:muonrec}, where 
  \PYTHIA-generated events are overlaid on minimum-bias
  \PbPb\ data.
The overlay is done independently of the \psin\ angles and, thus, should yield
  a zero \vn\ value when the analysis procedure used in the data is applied to 
  the simulated events.
Any systematic deviation from $\vn = 0$ seen in the simulated data is, then, a result of 
 jet bias. The procedure used to evaluate the jet bias in
 \vn\ values is as follows.

The presence of the recoil jet biases the observed $q$-vector in the FCal as\footnote{
  In this section, the two-dimensional $q$-vector is represented using complex numbers~\cite{HION-2015-14}.
 }:
\begin{eqnarray*}
{q}_{n}^{\mathrm{Biased}}=q_ne^{in\psin}+ ke^{in\phi^{\mathrm{Jet}}},
\end{eqnarray*}
where the first term on the right is the unbiased $q$-vector, which only 
  has the natural statistical smearing.
The second term on the right is the bias introduced by the recoil jet, 
  which shifts the event-plane angle to be aligned with the 
  recoil jet direction. 
The factor $k$ represents the strength of the bias and may depend on 
  the \pT\ of the recoil jet as well as the centrality, 
  and $\phi^{\mathrm{Jet}}$ is the direction of the jet.
Since the recoil jet is nominally back-to-back with the muon, its direction can 
  be written as:
\begin{eqnarray*}
 \phi^{\mathrm{Jet}}=\phi^{\mu}+\pi+\delta,
\end{eqnarray*}
where $\phi^{\mu}$ is the azimuthal angle of the muon and $\delta$
  represents event-by-event fluctuations in the jet direction.
This bias affects the numerator in the SP method (Eq.~\eqref{vn_sp});
the resolution (denominator in Eq.~\eqref{vn_sp}) is not affected by 
  the bias, as the resolution is calculated using minimum-bias events 
  and not from events that are triggered by muons.
The dot product between the muon's transverse direction and the biased 
  $q$-vector, averaged over many events, (numerator of Eq.~\eqref{vn_sp}) 
  now becomes 
\begin{eqnarray}
\langle e^{in\phi^{\mu}} (q_ne^{-in\psin}+ ke^{-in(\phi^{\mu}+\pi+\delta)})\rangle_{\mathrm{evts}}\
=\langle e^{in\phi^{\mu}} q_ne^{-in\psin} \rangle_{\mathrm{evts}} + \langle ke^{-in(\pi+\delta)} \rangle_{\mathrm{evts}}
\label{eq:vn_mc_correction}.
\end{eqnarray}
The first term on the right is the numerator of Eq.~\eqref{vn_sp} 
  for no bias, and the second term is the bias, which conveniently separates out
  as an additive contribution.
The second term on the right of Eq.~\eqref{eq:vn_mc_correction} corrected by 
  $\mathrm{Res}^{\mathrm{SP}}\{n\psin\}$ is the jet bias in $\vnsp$.

The bias determined in this manner is independent of \pt\
  within statistical errors.
The magnitude of the bias varies with centrality. 
It is smallest in the most central events -- where the underlying event is 
  quite large, and the additional energy deposited by the jet
  does not cause a significant perturbation -- and increases with 
  decreasing centrality.
For \vtwo, the \pt-averaged value of this bias is 0.0025 in the 0--10\% 
  centrality interval; it increases to 0.011 in the 40--60\% centrality 
  interval.
For comparison, the \vtwo at \pt = 4~\GeV\ in the 0--10\% and 40--60\% centrality intervals 
  is about 0.04 and 0.07, respectively.
Because the jet yield is suppressed by as much as a factor of two in
  \PbPb\ collisions~\cite{HION-2013-06}, only half of this estimated bias is applied as a
  correction.
Half of this estimated bias is also conservatively taken as the systematic 
  uncertainty of the correction.
In principle, the jet bias also affects the \RAA\ measurements since the
  correlated jet, if it falls within the FCal acceptance, also alters the centrality 
  interval to which the event is assigned.
However, this effect, estimated from the simulated-data overlay sample, is negligible 
  compared to the systematic uncertainties in the \RAA\ measurement 
  (Section~\ref{sec:raa_systematics}), and corrections for it are not applied.

\section{Systematic uncertainties}
\label{sec:systematics}
\subsection{Yield, cross-section, and \RAA\ systematic uncertainties}\label{sec:raa_systematics}

The measurements of the heavy-flavor muon differential cross-sections and
  per-event yields are subject to systematic uncertainties arising from
  the muon-trigger selection, muon-reconstruction efficiencies, the
  template-fitting procedure, muon \pT\ resolution, and the \pp\ luminosity.
They are described below.
Where appropriate, the uncertainties are smoothed as a function of \pt, 
  to reduce the statistical fluctuations in the uncertainty estimates.
The systematic uncertainties for the \PbPb\ data do not show any 
  significant variation with collision centrality.

The systematic uncertainty in the \PbPb\ muon-trigger efficiency is
  evaluated by varying the selections applied to the
  offline-reconstructed muons in the minimum-bias reference sample and
  re-evaluating the trigger efficiency. 
The resulting changes in the trigger efficiency are less than 0.5\% over $4 < \pT <
  14$~\GeV\ and are taken as the estimate of the systematic uncertainty
  in $\varepsilon_{\mathrm{trig}}$. 
The uncertainty in the \pp\ muon-trigger efficiency is evaluated similarly,
  and is less than 2.5\% for \pt\ $<$ 6~\GeV\ and less than 1.5\% for \pt\ $>$ 6~\GeV.
The systematic uncertainty associated with the muon-reconstruction 
  efficiency is evaluated by varying the muon selections,
  evaluating the reconstruction efficiency for the new selections,
  and repeating the analysis with the updated muon selection and reconstruction efficiency. 
This uncertainty is less than about 4\% for the \pp\ data and less than about 2.5\% for the \PbPb\ data. 
Separately, the minimum \pMSt\ (default value of 1.2~\GeV,  
  Section\,\ref{sec:muonrec}) is varied from 0.5~\GeV\ to 1.5~\GeV, and the entire 
  analysis is repeated. 
This variation affects the template fitting but also is sensitive to 
  potential systematic uncertainties in the muon reconstruction and trigger efficiencies. 
The change in the \PbPb\ muon yields from varying the minimum \pMSt, 
  taken as a systematic uncertainty in the heavy-flavor muon yields, 
  decreases with \pT\ from ${\sim} 10\%$ to ${\sim} 1.5\%$ 
  over the measured \pT\ range. 
For the \pp\ cross-section measurements, the systematic uncertainty decreases with \pT\ from 
  ${\sim} 11.5\%$ to ${\sim} 3\%$.
The systematic uncertainty associated with the \pMSt\ criterion is somewhat
  correlated with the systematic uncertainty associated with the trigger efficiency; 
  however, they are conservatively treated as independent uncertainties.

Systematic uncertainties resulting from the construction of the 
  templates, particularly the background template, are evaluated 
  by changing the relative proportions of different background 
  contributions. 
The pion and kaon decay-in-flight components of the background are 
  separately increased by a factor of two and then separately 
  decreased by a factor of two, as motivated by 
  differences observed in the kaon to pion yields 
  between \PYTHIA\ -- which is used to generate the MC templates -- 
  and data~\cite{Adam:2015qaa}.
For each variation, the template fitting is performed, and a new value for
  \fpr\ is obtained. 
The average of the unsigned differences between the varied and nominal 
  \fpr\ values is taken as the systematic uncertainty in the template 
  fitting due to the background composition. 
This is less than $0.5\%$ over the \pT\ range of the 
  measurement for both the \PbPb\ and \pp\ data. 

\newcommand{\efftrue}{\varepsilon_{\mathrm{true}}}
\newcommand{\efffake}{\varepsilon_{\mathrm{bkg}}}

In order to account for possible inconsistencies between the data and
MC templates that may arise from the effect of the trigger, or other
factors that may not be properly accounted for in the MC simulation, a
separate systematic uncertainty in the template-fitting method is
estimated using a ``cut-and-correct'' procedure applied to the \dpop\
distributions. In this procedure, the fraction of muons having $\Delta
p/p_\textrm{ID} < \dpopcut$, $f^<$, is measured in the
data in each centrality and \pT\ interval.
This fraction provides an estimate of the signal muon fraction, but it must 
be corrected for true muons having $\Delta p/p_\textrm{ID}
> \dpopcut$ (inefficiency) and background muons having $\Delta
p/p_\textrm{ID} < \dpopcut$ (fakes). The corrections are obtained from the MC signal
and background \dpop\ distributions and are expressed in terms of the
efficiencies,  $\efftrue$ and $\efffake$, for true and background muons, respectively, to
pass the $p/p_\textrm{ID} < \dpopcut$.
In terms of these efficiencies, $f^<$ is given by
\begin{eqnarray*}
f^< = f^{\mathrm{sig}}\efftrue + \left(1 - f^{\mathrm{sig}} \right) \efffake.
\end{eqnarray*}
Inverting this equation, the signal fraction estimated using the
cut-and-correct procedure is 
\begin{eqnarray*}
f^{\mathrm{sig}} = \dfrac{f^< - \efffake}{\efftrue - \efffake}.
\end{eqnarray*}
If the MC exactly describes the signal and background \dpop\
distributions in the data, then the cut-and-correct $f^{\mathrm{sig}}$
values will be identical
to the signal fractions obtained from the template fitting. Differences
from the template-fit signal fractions quantify the impact of
inaccuracies in the MC templates and are taken as a systematic
uncertainty. The cut-and-correct $f^{\mathrm{sig}}$ values
were evaluated using $\dpopcut = 0.1$. The obtained signal fractions were found
to be systematically higher than the results from the template fits at
both low and high \pT and in both the \pp\ and \PbPb\ data. The
relative difference is largest in the lowest \pT\ interval where it is ${\sim} 11\%$ and 
  6\% for the \pp\ and \PbPb\ data, respectively. 
It decreases with increasing \pt, and for the highest \pT\ interval, is 
  ${\sim}$6\% and 3\% for the \pp\ and \PbPb\ data, respectively. 

The \pp\ cross-sections and \PbPb\ per-event yields are not corrected 
  for any bin migrations that result from the muon momentum resolution. 
An evaluation of MC bin-by-bin correction factors gives values that 
  are typically within 1\% (2\%) of unity for \pp\ (\PbPb) data. 
These corrections are sufficiently small that they are not applied to the data. 
However, the deviations from unity are included in the systematic 
  uncertainties of the cross-sections and per-event yields. 

The measured \pp\ cross-section has an additional normalization
  systematic uncertainty of 3.1\% due to uncertainties in the 
  integrated luminosity. 

For the \RAA\ measurement, the systematic uncertainties from the 
  \pp\ cross-section and \PbPb\ per-event yields are propagated 
  as if they are correlated, i.e.\ the systematic variations 
  are simultaneously performed in the \pp\ and \PbPb\ data and the change 
  in the \RAA\ value is taken as the systematic uncertainty.
Besides the systematic uncertainties from the \pp\ cross-section and \PbPb\ per-event yields, 
  additional systematic uncertainties in the \RAA\ measurement come from theoretical uncertainties 
  in \ATAA, which are listed in Table~\ref{tab:glauber}.
Table~\ref{tab:RAA_syst_errors} summarizes the final experimental systematic 
  uncertainties in \RAA. 
The total uncertainty is obtained by adding the individual 
  uncertainties in quadrature.

\begin{table}
\caption{
Relative systematic uncertainties in the heavy-flavor muon \RAA,
  quoted in percent, for selected \pt\ intervals. 
\label{tab:RAA_syst_errors}}
\begin{centering}
\begin{tabular}{|l|S[table-format=2.1] S[table-format=2.1] S[table-format=2.1]| c}\cline{1-4} 
  \pt interval                      &{$4<\pt<4.5$~\GeV}   & {$6<\pt<7$~\GeV}   & {$10<\pt< 12$~\GeV}&\tl\cline{1-4}
  Muon selection [\%]               &  2.5                &   4                & 4                  &\tl
  \pMSt\ selection [\%]             &  7.5                &   2                & 2                  &\tl
  Background template variation [\%]&  0.5                &   0.5              & 0.5                &\tl
  Template fitting [\%]             &  13                 &   7                & 5                  &\tl
  Efficiency  [\%]                  &  2.5                &   1.5              & 1.5                &\tl\cline{1-4}
\end{tabular}
\par\end{centering}
\end{table}

\subsection{Systematic uncertainties in \vn}\label{sec:vn_systematics}

The sources of the systematic uncertainties in the \vn\ measurements are 
  primarily the same as those in the \RAA\ measurements (Section~\ref{sec:raa_systematics}). 
However, several sources of systematic uncertainty that affect \RAA\ do not have a significant 
  effect on the \vn\ values. 
The \vn\ measurements are independent of the trigger and tracking efficiencies. 
While these efficiencies have an impact on the absolute muon yields, the 
  \vn\ values, which measure the relative or fractional modulation 
  in yields, are insensitive to them. 
Therefore, the uncertainties in the efficiencies do not have any effect 
  on the \vn\ measurements. 
Varying the muon selection as described in Section~\ref{sec:raa_systematics}
  changes the measured value of \vtwo\ by (1--2)$\times 10^{-3}$ below \pt\ of 6~\GeV.
The \pMSt\ criterion variation changes the measured value of \vtwo\ by (0.5--1)$\times10^{-3}$
  for $\pt < 6$~\GeV.
At higher \pt\ the effect of this criterion on \vtwo\ is about $0.2\times10^{-3}$.
For \vthree\ and \vfour\ the effect of the \pMSt\ criterion is (0.5--1)$\times10^{-3}$
  across the measured \pt\ range.
The effects of the muon selection and the \pMSt\ criterion are  evaluated 
  not just by applying the selection in the data but also by
  rebuilding the templates in the MC simulation while applying the variations, and then 
  repeating the entire analysis. 
The variation in the shape of the background template, when varying the 
  relative contribution of the pion and kaon backgrounds, 
  results in variations in the \vn\ values 
  that are less than $0.5\times10^{-3}$ across most of the 
  centrality and \pt\ ranges.
The systematic uncertainty in \vn\ due to \pt-resolution effects 
  is estimated to be less than 1\% (relative) for $\pt < 10$~\GeV.
This estimate is obtained by first determining the \pt\ resolution 
  using MC simulation (Section~\ref{sec:muonrec}), and then evaluating the change 
  in the \vtwo\ values when smearing the \pt\ of the reconstructed muons 
  by this resolution.
The uncertainty arising from the \pt\ resolution is treated as a fractional
  uncertainty; since if \vn\ changes, then the \pt\ resolution effects
  that result in migration of muons from one \pt\ interval to an adjacent one
  also increase proportionally.
For $\pt > 10$~\GeV, the systematic uncertainties from all the above 
  sources are partially correlated with the statistical uncertainties, 
  and are thus somewhat larger.

Additional systematic uncertainties that affect only the \vn\ but not 
  the \RAA\ measurements are the uncertainty in the EP resolution for 
\psin\ and the jet bias correction discussed in Section~\ref{sec:vtwo_method}.
The uncertainty in the EP resolution is a relative uncertainty and depends 
  only on the centrality. 
It varies between 1\% and 5.5\% depending on the harmonic and centrality.
The systematic uncertainty associated with the jet bias correction 
  is the leading uncertainty in the measurement.
The absolute value of this uncertainty depends on the centrality 
  and the harmonic order but is independent of \pt.
It increases monotonically from central to peripheral events and
  is much larger for \vthree\ and \vfour\ than for \vtwo.
Table~\ref{tab:v4_syst_errors} summarizes 
  the systematic uncertainties for the \vn\ in three different \pt\ ranges 
  and for two centrality intervals.
The uncertainties associated with the \pt\ resolution and EP resolution
  are intrinsically fractional uncertainties and are listed as percentages.
All other uncertainties are listed as absolute values.

\begin{table}
\caption{
Systematic uncertainties in the heavy-flavor muon \vn\ 
  for selected \pt\ and centrality intervals. 
The values are for the EP method and are quoted either as absolute values or in percent.
They are averaged over \pt\ intervals that are larger than the 
  intervals used for the measurement.
\label{tab:v4_syst_errors}}
\begin{centering}
\begin{tabular}{|c|l| S[table-format=2.1]  S[table-format=2.1]  S[table-format=2.1]  S[table-format=2.1]  S[table-format=2.1]  S[table-format=2.1]| c}\cline{1-8} 
&\pt\ interval             &\multicolumn{2}{c}{$4<\pt<5$~\GeV}&\multicolumn{2}{c}{$6<\pt<10$~\GeV} &\multicolumn{2}{c|}{$10<\pt<14$~\GeV}\tl
&Centrality                               &\text{0--10\%}&\text{40--60\%}  &\text{0--10\%}&\text{40--60\%} &\text{0--10\%} &\text{40--60\%}\tl\cline{1-8}
{\multirow{7}{*}{\vtwo}}&
$\pMSt$ selection              [$10^{-3}$]&  0.6         &     1.0         &   0.2        &    0.3              &0.2      &0.3      &\tl 
&Muon selection                [$10^{-3}$]&  1.0         &     1.2         &   2.0        &    3.0              &2.0      &3.0      &\tl
&Background template variation [$10^{-3}$]&  0.1         &     0.5         &   0.1        &    0.5              &0.1      &0.5      &\tl 
&Template fitting              [$10^{-3}$]&  0.1         &     0.1         &   0.1        &    0.1              &0.1      &0.1      &\tl
&Jet bias correction           [$10^{-3}$]&  1.2         &     5.5         &   1.2        &    5.5              &1.2      &5.5      &\tl 
&\pt\ resolution [\%]                     &  1.0         &     1.0         &   1.0        &    0.4              &0.6      &0.6      &\tl 
&EP resolution [\%]                       &  3.7         &     3.3         &   3.7        &    3.3              &3.7      &3.3      &\tl\cline{1-8} 
{\multirow{7}{*}{\vthree}}&
$\pMSt$ selection [$10^{-3}$]             &  0.3         &     0.2         &   0.3        &    0.2              &0.3      & 0.2     &\tl 
&Muon selection  [$10^{-3}$]              &  0.8         &     3.0         &   0.8        &    3.0              &0.8      & 3.0     &\tl
&Background template variation [$10^{-3}$]&  0.5         &     0.5         &   0.5        &    0.5              &0.5      & 0.5     &\tl 
&Template fitting [$10^{-3}$]             &  0.1         &     0.1         &   0.1        &    0.1              &0.1      & 0.1     &\tl
&Jet bias correction  [$10^{-3}$]         &  1.7         &    11.0         &   1.7        &   11.0              &1.7      &11.0     &\tl 
&\pt\ resolution [\%]                     &  1           &     1           &   1          &    1                &1        & 1       &\tl 
&EP resolution [\%]                       &  3.3         &     5.4         &   3.3        &    5.4              &3.3      & 5.4     &\tl\cline{1-8} 
{\multirow{7}{*}{\vfour}}&
$\pMSt$ selection [$10^{-3}$]             &  0.5         &    0.8          &   0.5        &    0.8            & 0.5       & 0.8       &\tl 
&Muon selection [$10^{-3}$]               &  0.8         &    0.6          &   0.8        &    0.6            & 2.0       & 2.0      &\tl
&Background template variation [$10^{-3}$]&  0.2         &    0.5          &   0.2        &    0.5            & 0.2       & 1.5      &\tl 
&Template fitting [$10^{-3}$]             &  0.1         &    0.1          &   0.1        &    0.1            & 0.1       & 0.1      &\tl
&Jet bias correction  [$10^{-3}$]         &  1.8         &    15           &   1.8        &   15              & 1.8       & 15       &\tl 
&\pt\ resolution [\%]                     &  1           &    1.0          &   1.0        &   1.0             & 1.0       & 1.0      &\tl 
&EP resolution [\%]                       &  4.1         &    5            &   4.1        &   5               & 4.1       & 5        &\tl\cline{1-8} 
\end{tabular}
\par\end{centering}
\end{table}

\FloatBarrier
\section{Results}
\label{sec:results}
\subsection{Heavy-flavor muon \RAA}
\label{sec:RAA}

\begin{figure}[!b]
\centerline{
\includegraphics[width=0.9\columnwidth]{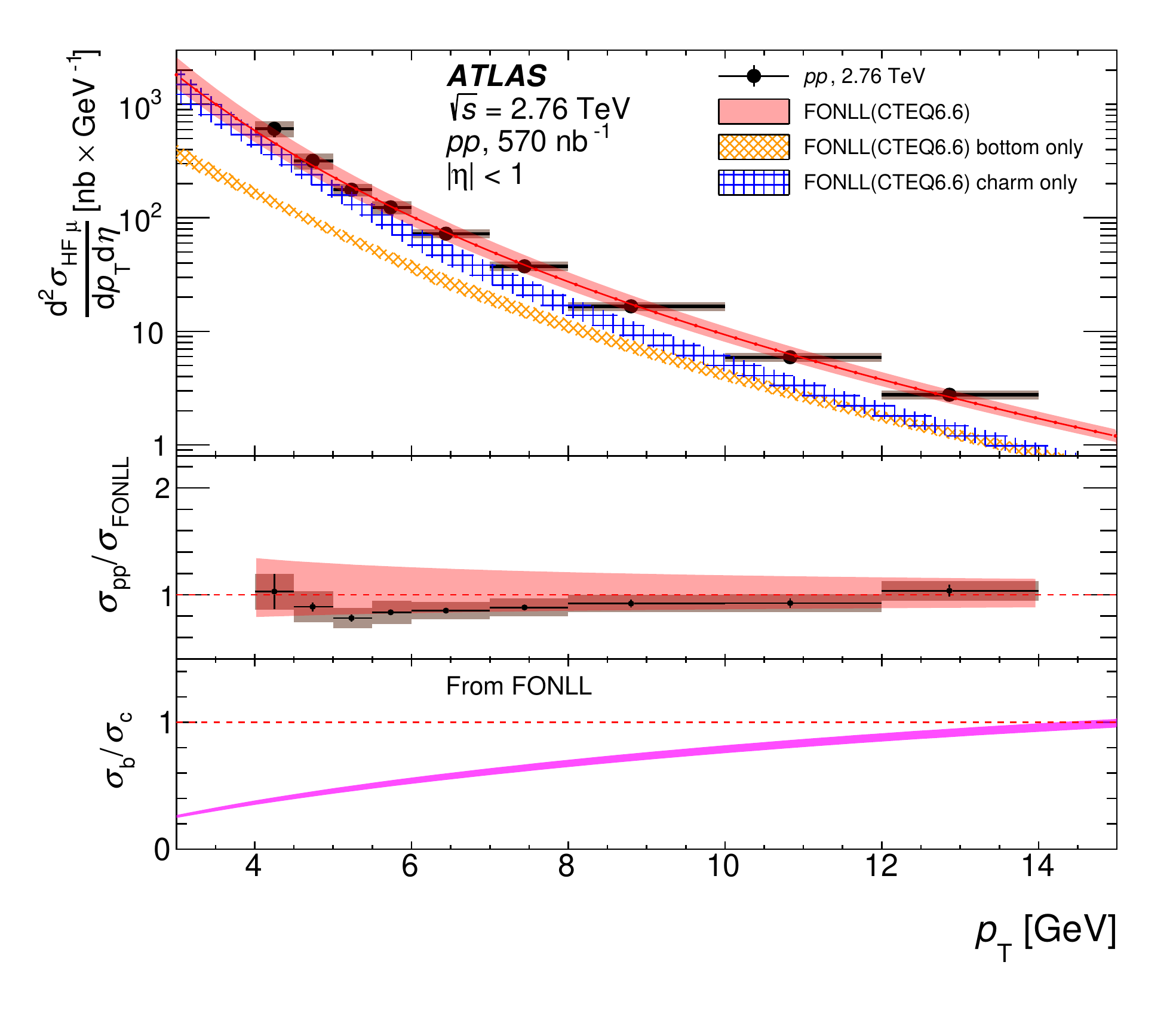}
}
\caption{
Top panel: the \pT\ dependence of the measured  heavy-flavor muon 
  cross-section in $\sqrt{s}=2.76$~\TeV\ $pp$ collisions. The data
  points are plotted at the average muon \pT\ within a given \pT\ interval.
The vertical bars and bands on the data points indicate statistical
  and systematic uncertainties, respectively.
The cross-section for heavy-flavor decays from FONLL
  calculations is also shown, along with the individual contributions
  from bottom and charm quarks.
For the FONLL calculations, the vertical width of the band 
  represents theoretical systematic uncertainties.
Middle panel: the ratio of the measured and FONLL cross-sections
  integrated over each \pT\ interval. 
Statistical and systematic uncertainties in the data are indicated 
  by error bars and gray shaded boxes, respectively. 
The systematic uncertainty of the ratio from FONLL is indicated 
  by the shaded band centered on unity.
Bottom panel: the ratio of the bottom contribution to the charm 
  contribution in the FONLL calculations.
All results are averaged over $|\eta|<1$.
} 
\label{fig:FONLL}
\end{figure}

Figure~\ref{fig:FONLL} shows the measured heavy-flavor muon cross-sections, 
  calculated via Eq.~\eqref{eq:ppsig}, in the \sqs\ = 2.76~\TeV\ $pp$ 
  data as a function of the muon \pt. 
The error bars show statistical uncertainties resulting from combining 
  the statistical uncertainties of $\Delta N_{\mu}$ and \fpr. 
The measured cross-sections are compared with
  fixed-order plus next-to-leading-logarithm (FONLL)~\cite{Cacciari:1998it,Cacciari:2001td,Cacciari:2012ny,Cacciari:2015fta}
  calculations using CTEQ~6.6 PDFs~\cite{Nadolsky:2008zw}.
The FONLL calculations are based on three main components: 
  1) the heavy-quark production cross-section calculated 
     in perturbative QCD by matching the fixed-order next-to-leading-order (NLO) terms with the
     next-to-leading-logarithms (NLL) high-\pT\ resummation, 
  2) the non-perturbative heavy-flavor fragmentation functions determined from $e^+e^-$
     collisions and extracted in the same framework, and 
  3) the decays of the heavy hadrons 
     to leptons using decay tables and form factors from $B$-factories.
The middle panel of Figure~\ref{fig:FONLL} presents the ratios of the
  measured and FONLL cross-sections.
The FONLL calculation agrees with the data within systematic uncertainties. 
The individual contributions of the bottom and charm quarks to the 
  heavy-flavor muon cross-section obtained from the FONLL calculations are compared in the 
  lower panel of Figure~\ref{fig:FONLL}.
It is seen that at 4~\GeV\ the contribution of the bottom quark to the
  muon cross-section is about 40\% of that of the charm quark. 
The relative contribution increases monotonically with the muon \pt,
  and at \pt = 14~\GeV, the contributions from
  bottom and charm decays are comparable.

\begin{figure}[!b]
\centerline{
\includegraphics[width=0.9\columnwidth]{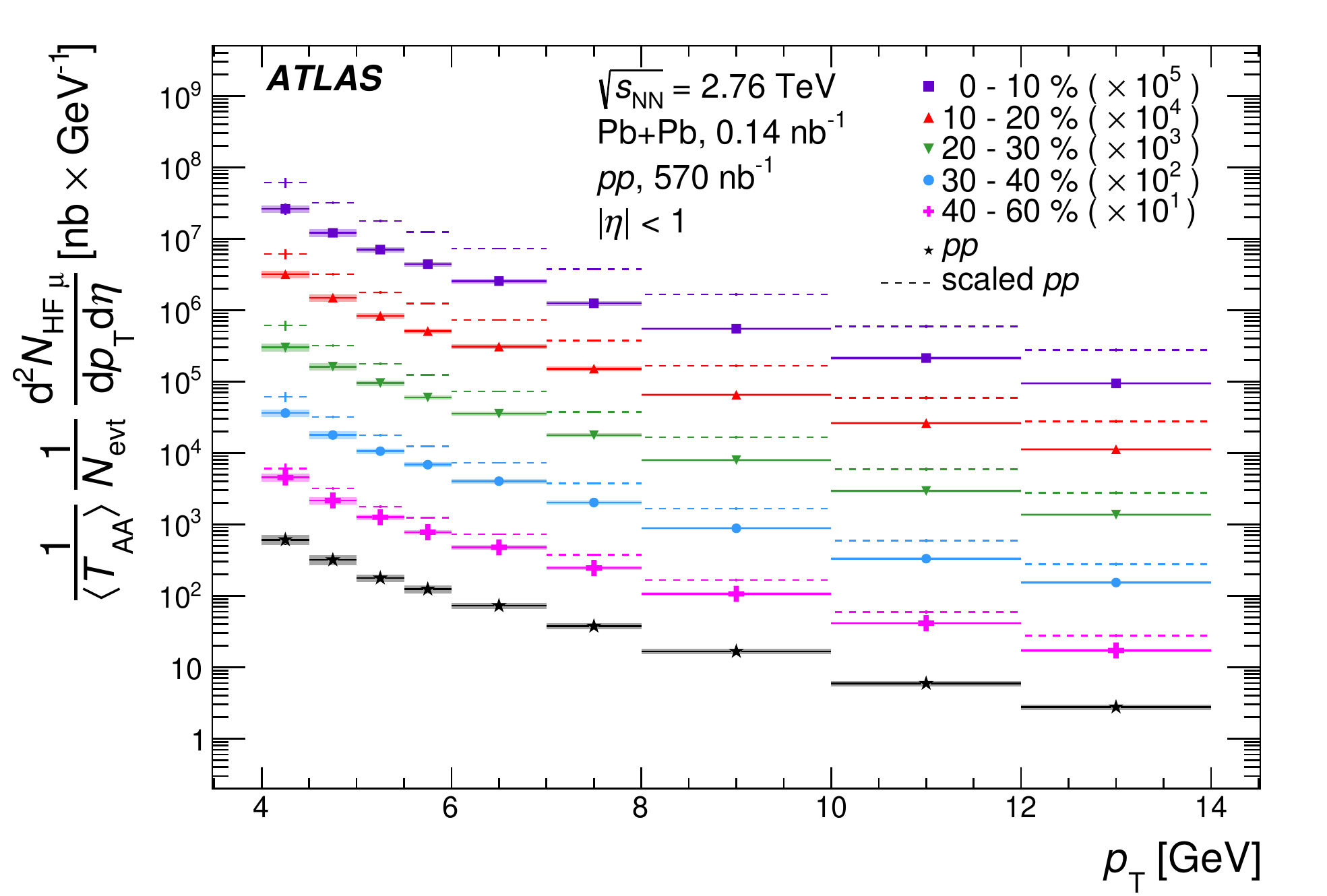}
}
\caption{
The \pT\ dependence of the measured \PbPb\ heavy-flavor muon 
  differential per-event yields for different centrality intervals
  scaled by the corresponding \ATAA.
Also shown is the measured \pp\ heavy-flavor muon differential 
  cross-section.
For clarity, the results for the different centralities are multiplied by 
  scale factors that are indicated in the legend.
The \pp\ cross-section is re-plotted multiple times, as dashed lines, 
  multiplied by these scale factors,
  for comparison with the results for the different Pb+Pb centralities.
The error bars and shaded bands represent statistical and systematic 
  uncertainties, respectively, and in many cases are too small 
  to be seen.
} 
\label{fig:yieldPbPb}
\end{figure}

Figure~\ref{fig:yieldPbPb} shows the differential per-event heavy-flavor muon
  yields in \PbPb\ collisions (Eq.~\eqref{eq:yieldPbPb}) scaled by 
  the corresponding \ATAA\ for the centrality intervals in this analysis.
The  statistical uncertainties are the combined statistical uncertainties 
  of $\Delta N_{\mu}$ and  \fpr.
Figure~\ref{fig:yieldPbPb} also compares the \ATAA\ scaled yields to the measured 
  \pp\ cross-section.
There are significant differences between the scaled \PbPb\ yields and the 
  \pp\ cross-section, which monotonically increase with 
  increasing centrality.

\begin{figure}[!h]
\centerline{
\includegraphics[width=1.0\columnwidth]{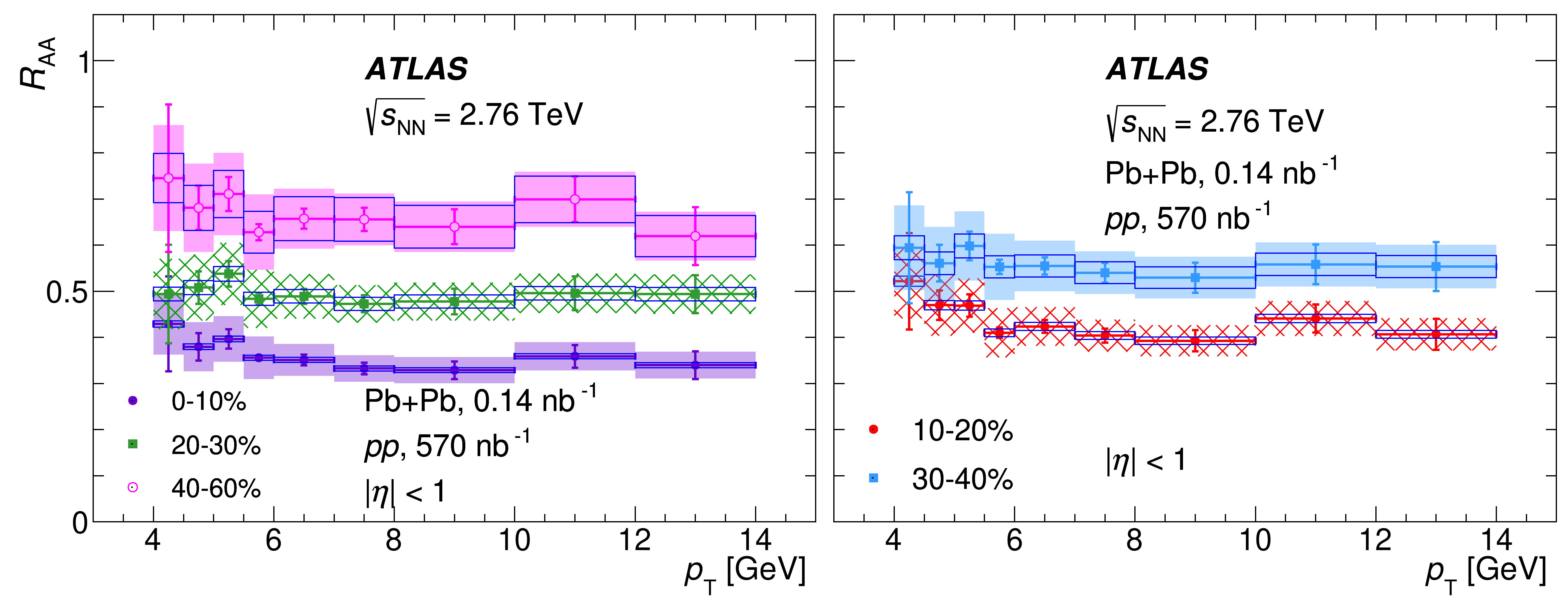}
}
\caption{
The measured \PbPb\ heavy-flavor muon \RAA\ as a function of \pT. 
For clarity, the centrality intervals are split between the two panels.
The left panel shows results for the
  0--10\%, 20--30\%, and 40--60\% centrality intervals while the right
  panel shows results for the 10--20\% and 30--40\% intervals.
The error bars represent statistical uncertainties.
The boxes indicate theoretical uncertainties of \ATAA.
The shaded bands represent the experimental systematic uncertainties.
} 
\label{fig:raa}
\end{figure}

The heavy-flavor muon \RAA\ is calculated according to Eq.~\eqref{eqn:raa_def}
  using the results in Figure~\ref{fig:yieldPbPb} and is shown in Figure~\ref{fig:raa}.
The parameter \RAA\ does not depend on \pT\ within the uncertainties of the measurement. 
This is of note because the suppression of bottom and charm quarks in the quark--gluon plasma (QGP) 
  is expected to be different, and the FONLL calculations show that the contribution 
  of bottom and charm quarks changes with \pt\ in the \pp\ case, as shown in Figure~\ref{fig:FONLL}.
The parameter \RAA\ decreases between peripheral 40--60\% collisions, where it is about 0.65,
  to more central collisions, reaching a value of about 0.35 in the 0--10\% 
  centrality interval. 

\begin{figure}[!h]
\centerline{
\includegraphics[width=1.0\columnwidth]{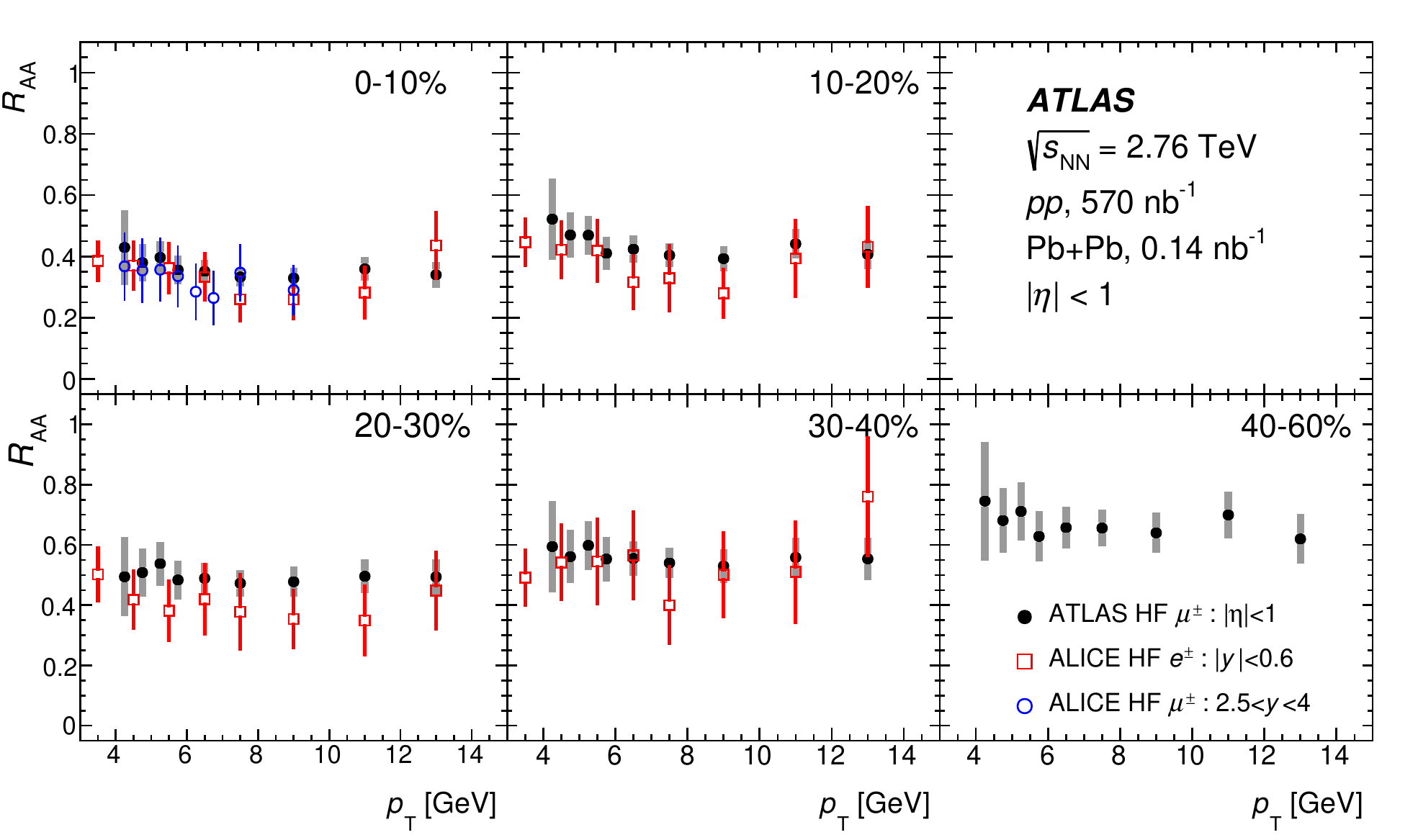}
}
\caption{
Comparison of the \PbPb\ heavy-flavor muon \RAA\ measured in this analysis to
  similar measurements for muons at forward rapidity ($2.5 < y < 4$) 
  and heavy-flavor electrons at mid-rapidity ($|y| < 0.6$)
  from the ALICE Collaboration.
The error bars represent systematic 
  and statistical uncertainties added in quadrature.
The \ATAA\ errors are identical between the three measurements 
  and are excluded from the comparison.
}
\label{fig:compare_alice}
\end{figure}

Figure~\ref{fig:compare_alice} shows a comparison of the \RAA\ 
  measurements in this \papertype\ with similar measurements for muons 
  at forward rapidity (2.5$ < y < $4)~\cite{Abelev:2012qh}
  and heavy-flavor electrons at mid-rapidity ($|y| < 0.6$)~\cite{Adam:2016khe} from 
  the ALICE Collaboration.
In general, the results are consistent; however, the present 
  measurements have considerably smaller uncertainties.

\begin{figure}[!h]
\centerline{
\includegraphics[width=1.0\columnwidth]{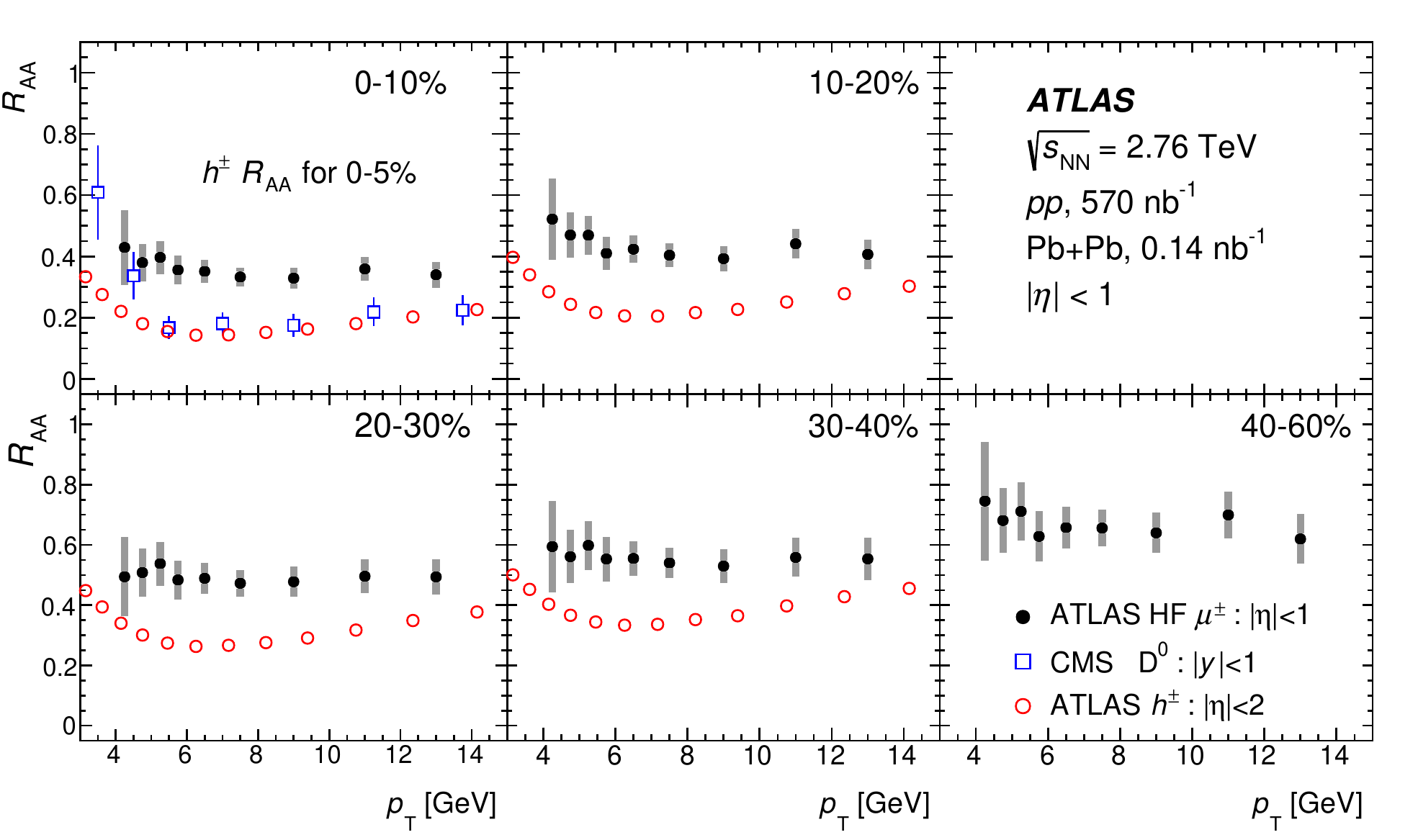}
}
\caption{
Comparison of the \PbPb\ heavy-flavor muon \RAA\ measured in this analysis to
  the \RAA\ for inclusive charged hadrons from ATLAS and
  the \RAA\ for identified $D^0$ mesons from the CMS Collaboration.
The error bars represent systematic 
  and statistical uncertainties added in quadrature.
The \ATAA\ errors are identical between the three measurements 
  and are excluded from the comparison.
The inclusive charged hadron  \RAA\ values shown in the top left panel 
  are for the 0--5\% centrality interval.
}
\label{fig:compare_D0}
\end{figure}

Figure~\ref{fig:compare_D0} compares the \RAA\ measurement presented in this paper 
  with the \RAA\ of inclusive charged hadrons~\cite{HION-2011-03} at $\sqn=2.76$~\TeV\ 
  and identified $D^{0}$ mesons~\cite{CMS-HIN-16-001} from the CMS Collaboration at $\sqn=$~5.02~\TeV.
The \RAA\ from $D^0$ analyses is similar to that of inclusive hadrons for $\pt > 5$~\GeV~\cite{CMS-HIN-16-001},
  implying that the charm suppression is very similar to  
  that for the light quarks and gluons.
On the other hand, the heavy-flavor muon \RAA, which includes contributions 
  from bottom and charm, is observed to be larger than that of inclusive hadrons.
This would imply a significantly smaller suppression for muons from the 
  decays of $b$-hadrons. 
One caveat is that the  $D^0$ \pt\ and 
  the HF muon \pt\ are related differently to the \pt\ of the HF quark that 
  produced them.
However, this effect is mitigated by the relatively weak \pt\ dependence of both
 the $D^0$ and HF muon \RAA\  over the 4--14~\GeV\ \pt\ range.

\FloatBarrier
\subsection{Heavy-flavor muon \vn}\label{sec:vtwo}

\begin{figure}[!h]
\centerline{
\includegraphics[width=0.99\columnwidth]{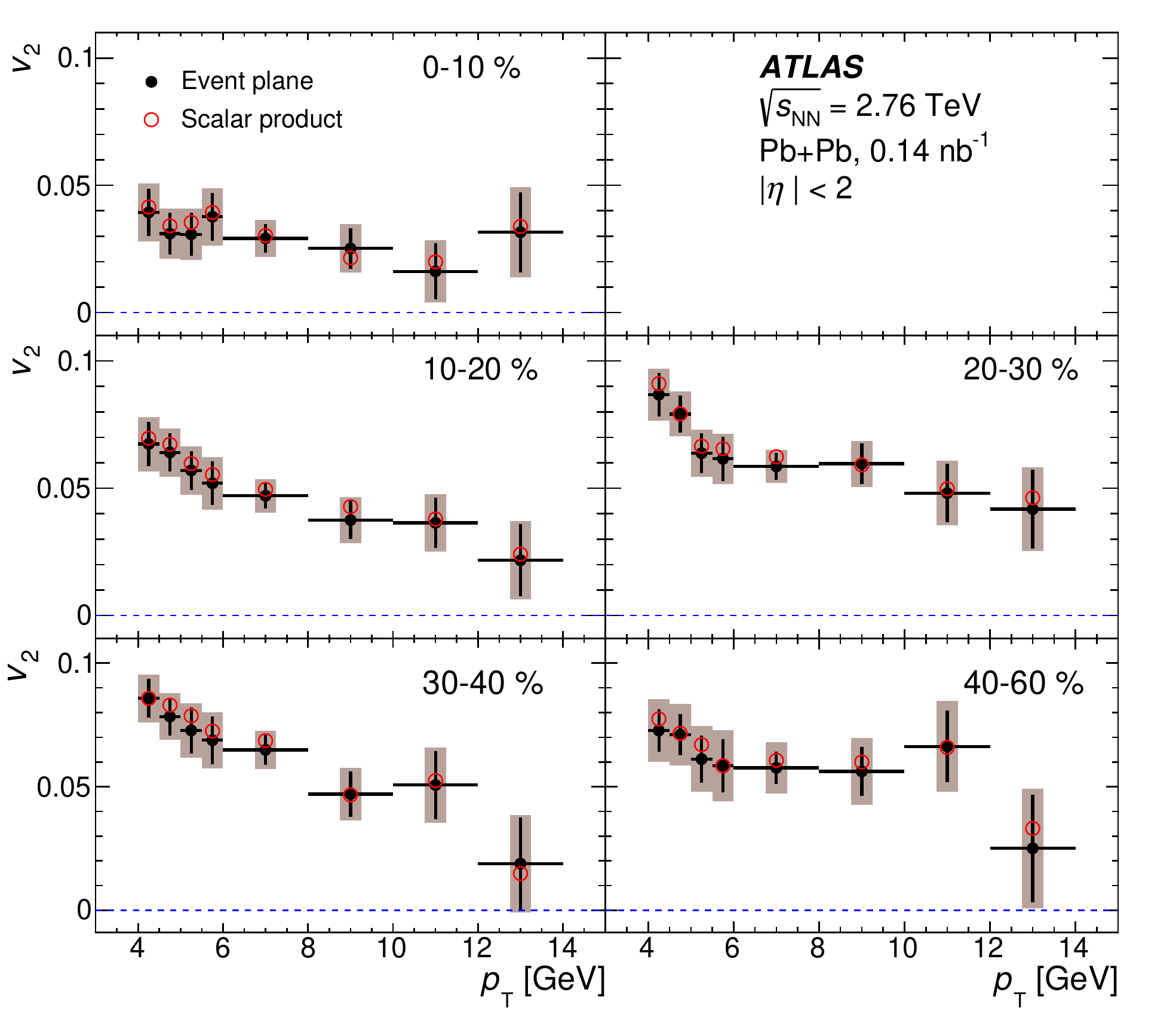}
}
\caption{
The \pt\ dependence of the \PbPb\ heavy-flavor muon \vtwo.
Results are shown for both the EP and SP methods.
Each panel represents a different centrality interval. 
The error bars and shaded bands represent statistical
  and total uncertainties, respectively, and are shown only for the EP $\vtwo$.
The horizontal dashed lines indicate $\vtwo = 0$. 
} 
\label{fig:v2_pt_dep}
\end{figure}

Figure~\ref{fig:v2_pt_dep} shows the \vtwo\ values measured using the EP method 
  as a function of \pt\ for the five centrality intervals in this analysis, 
  including the statistical and total uncertainties.
The evaluation of the total uncertainty includes the 
  correlation between the statistical uncertainties and the systematic 
  uncertainties that are proportional to \vn, 
  i.e.\ the relative uncertainties associated with the EP and \pt\ resolutions.
This correlation arises because as the measured \vn\ is varied within its 
  statistical uncertainty, the relative uncertainties that are proportional to 
  \vn\ also vary.
The other (absolute) systematic uncertainties are added in quadrature
  to the correlated uncertainty to get the total uncertainty.
Over the 10--40\% centrality range, $v_2$ is largest at the lowest measured \pt\ 
  of 4~\GeV\ and decreases for higher \pt.
However, in the 0--10\% and 40--60\% centrality intervals, 
  no clear \pt\ dependence is visible.
For all centralities, a significantly non-zero \vtwo\ is observed up to 
  a \pt\ of 12~\GeV. 
Figure~\ref{fig:v2_pt_dep} also shows the \vtwosp\ values, which 
  are slightly higher than the EP values.
The systematic uncertainties and a significant fraction of the statistical 
  uncertainties are correlated between the EP and SP \vtwo\ values, and for
  clarity are not shown for the $\vtwosp$.
These measurements are consistent with previous \vtwo\ measurements of 
  heavy-flavor muons~\cite{Adam:2015pga} and heavy-flavor electrons~\cite{Adam:2016ssk} from the ALICE Collaboration,
  but have significantly smaller statistical and systematic uncertainties,
  and are performed over wider centrality and \pt\ ranges.

\begin{figure}[!h]
\centerline{
\includegraphics[width=0.99\columnwidth]{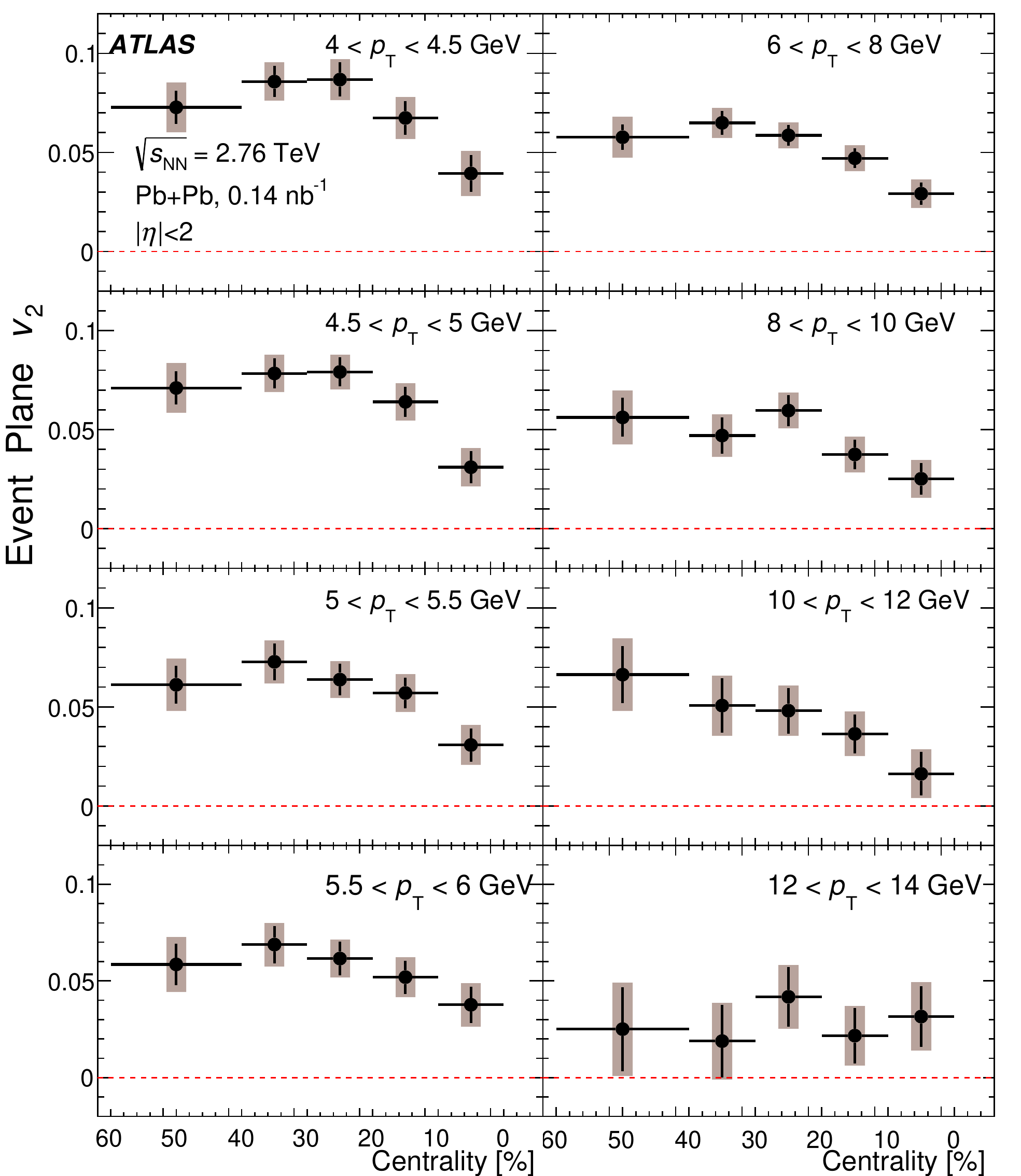}
}
\caption{
The centrality dependence of the \PbPb\ heavy-flavor muon \vtwo\
(the horizontal scale decreases in centrality). 
Each panel represents a different \pt\ interval. 
The error bars and shaded bands represent statistical
  and total uncertainties, respectively.
The dashed lines indicate $\vtwo = 0$.
The results are for the EP method.
} 
\label{fig:v2_cent_dep}
\end{figure}

Figure~\ref{fig:v2_cent_dep} shows the \vtwo\ obtained from the EP method plotted 
  as a function of centrality for different \pt\ intervals.
For \pt\ in the range 4--8~\GeV, the centrality dependences of the 
  heavy-flavor muon \vtwo\ are qualitatively similar in shape, but considerably 
  smaller in magnitude, to those for charged hadrons of similar 
  \pt~\cite{HION-2011-05,HION-2011-01}. 
In this \pt\ range, the \vtwo\ first increases from central to mid-central
  events, reaches a maximum between 20\% and 40\% centrality, and 
  then decreases. 
Over the \pt\ range of 8--12~\GeV, some deviation from this trend is 
  observed, with the \vtwo\ increasing monotonically from central 
  to peripheral events. 
However, the associated statistical and systematic uncertainties are 
  considerably larger.
This monotonically increasing centrality dependence of the \vtwo\ at high
  \pt\ is also seen in the inclusive charged hadron 
  \vtwo~\cite{HION-2011-05,HION-2011-01}.
For the highest \pt\ interval of $12 < \pt < 14$~\GeV, the statistical and 
  systematic errors are too large to identify a clear centrality 
  dependence of \vtwo.

\begin{figure}[!h]
\centerline{
\includegraphics[width=0.99\columnwidth]{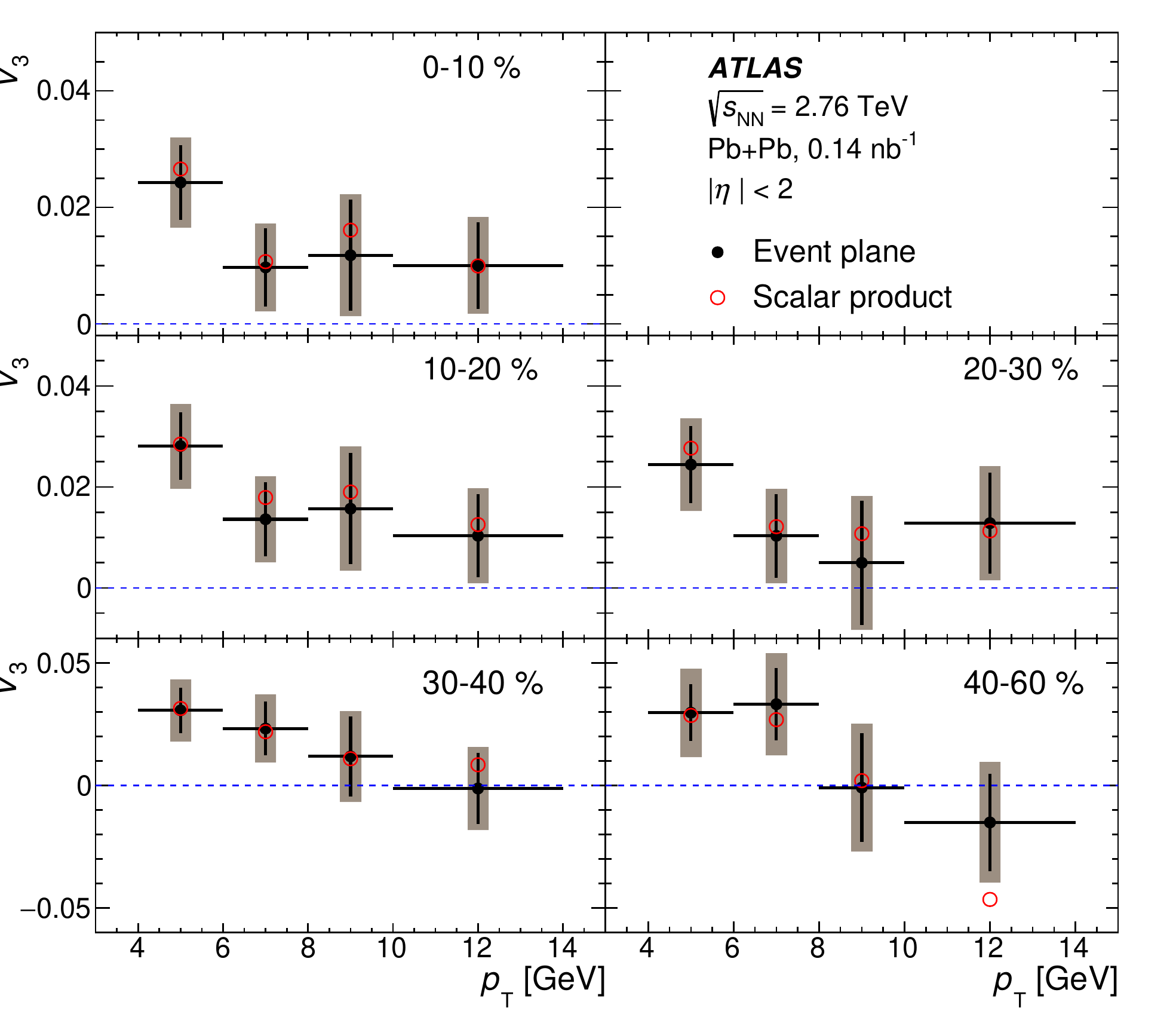}
}
\caption{
The \pt\ dependence of the \PbPb\ heavy-flavor muon $v_{3}$.
Results are shown for both the EP and SP methods.
Each panel represents a different centrality interval. 
The error bars and shaded bands represent statistical
  and total uncertainties, respectively, and are shown only for the EP $\vthree$.
The horizontal dashed lines indicate $v_{3}=0$. 
} 
\label{fig:v3_pt_dep}
\end{figure}

\begin{figure}[!h]
\centerline{
\includegraphics[width=0.99\columnwidth]{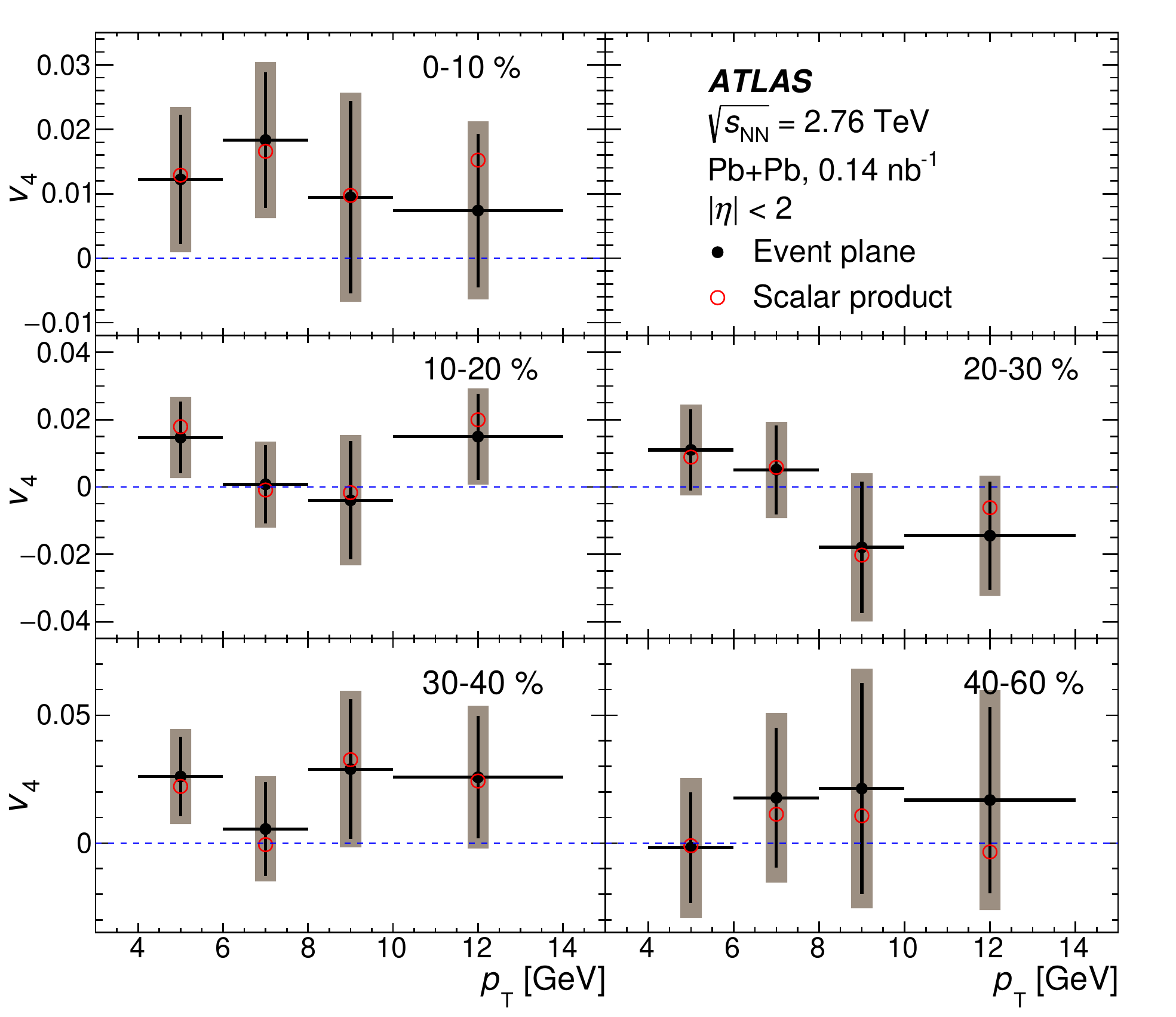}
}
\caption{
The \pt\ dependence of the \PbPb\ heavy-flavor muon $v_{4}$.
Results are shown for both the EP and SP methods.
Each panel represents a different centrality interval. 
The error bars and shaded bands represent statistical
  and total uncertainties, respectively, and are shown only for the EP $\vfour$.
The horizontal dashed lines indicate $v_{4}=0$. 
} 
\label{fig:v4_pt_dep}
\end{figure}

Figure~\ref{fig:v3_pt_dep} shows the \pt\ dependence of \vthree.
At a given \pt\ and centrality, \vthree\ is a factor of 2--3 smaller 
  than the corresponding \vtwo.
As with \vtwo, \vthree\ also decreases with increasing 
  \pt\ over the 4--8~\GeV\ \pt\ range.
At higher \pt, the statistical uncertainties are too large
  to observe clear \pt-dependent trends.
The parameter \vthree\ shows a much weaker variation with centrality: the \vthree\
  values at a given \pt\ are consistent within uncertainties across the different 
  centrality intervals.
These features for the centrality and \pt\ dependence are consistent 
  with observations of the inclusive 
  charged-hadron \vthree~\cite{HION-2011-01}.
Figure~\ref{fig:v4_pt_dep} shows the \pt\ dependence of \vfour.
The statistical uncertainties in \vfour\ 
 do not allow inference of any significant \pt- or 
 centrality-dependent trends.

\FloatBarrier
\subsection{Comparison with theoretical models}
\label{sec:theory_compare}
In this section, the measured \RAA\ and \vtwo\ values are compared with calculations
  from the TAMU transport model~\cite{He:2014cla} and 
  the DABMod model~\cite{Prado:2016szr}.
TAMU is a transport model for heavy flavor within the QGP and subsequent hadronic phase.
The initial heavy-quark spectra used in the model are obtained from FONLL calculations, 
  accounting for shadowing effects in Pb+Pb collisions.
The space-time evolution of the bulk QGP medium, in which the heavy quarks diffuse, 
  is modeled using ideal relativistic hydrodynamics, tuned to reproduce the charged-hadron \pT\ spectra 
  and inclusive elliptic flow measured in \PbPb\ collisions at the LHC.
The initial conditions for the hydrodynamic modeling are obtained from the Glauber 
  model and do not include initial state fluctuations or initial flow.
After this tuning, there are no free parameters in the model. 
The hadronization of heavy-flavor quarks is done partially via recombination 
  of heavy quarks with light-flavor hadrons in the QGP and partially by fragmentation.
Finally, the diffusion of heavy-flavor hadrons in the hadronic phase is 
  continued until kinetic freeze-out.
DABMod is an energy-loss model for heavy quarks traversing 
  the QGP.
The energy loss is a parameterized analytic function of the 
  velocity of the heavy quark and the local temperature.
The initial \pt\ distribution of heavy quarks is obtained from FONLL calculations.
The underlying QGP is modeled using (2+1)-dimensional relativistic viscous hydrodynamics 
  including event-by-event fluctuations in the initial conditions and 
  subsequent hydrodynamic expansion. 
All the hydrodynamic parameters are tuned to describe the 
  experimental flow data at low \pt.
The heavy quarks are evolved on top of the hydrodynamic underlying event until they 
  reach a decoupling temperature below which they are hadronized via fragmentation.
Any subsequent hadronic rescattering is neglected.
The DABMod model calculations are available for \RAA\ and \vtwo--\vfour 
  for all the centrality intervals over which the measurements are performed 
  in this \papertype.
The TAMU calculations for \RAA\ are available for the 0--10\%, 
  20--40\% and 40--60\% centrality intervals, and for \vtwo\ 
  for the  20--40\% and 40--60\% centrality intervals only.

\begin{figure}[!b]
\centerline{
\includegraphics[width=0.9\columnwidth]{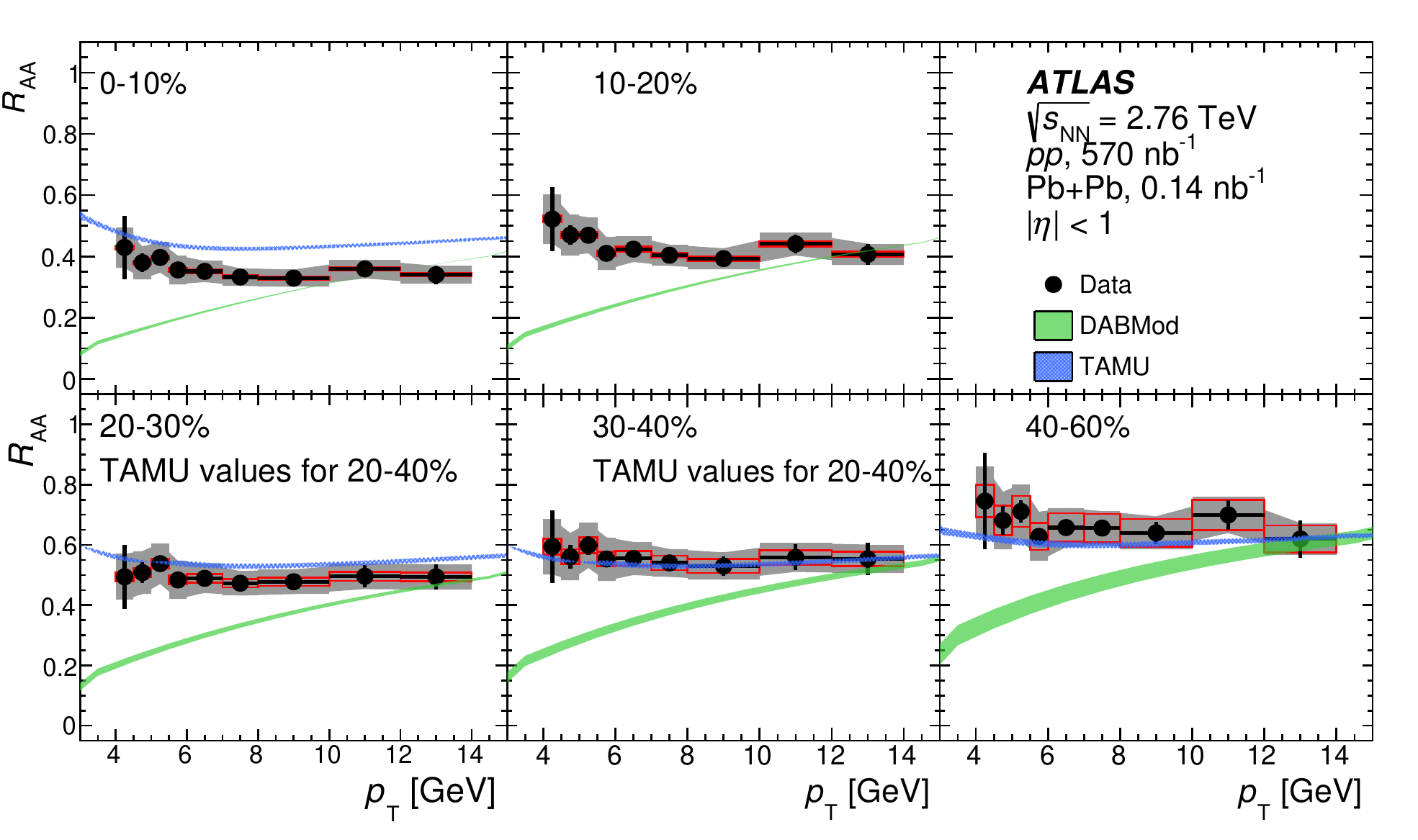}
}
\caption{
Comparison of the measured heavy-flavor muon \RAA\ in \PbPb\ collisions 
  with the values predicted from the TAMU transport model and the DABMod model.
Each panel represents a different centrality interval.
For the 20--30\% and 30--40\% centrality intervals, the plotted TAMU values correspond to
  the 20--40\% centrality interval.
For the data, the error bars represent statistical uncertainties,
  the shaded bands represent the experimental systematic uncertainties,
  and the boxes indicate theoretical uncertainties from \ATAA.
For the model calculations the bands indicate the theoretical 
  systematic uncertainties.
} 
\label{fig:RAA_theory}
\end{figure}

Figure~\ref{fig:RAA_theory} compares the measured heavy-flavor muon \RAA\ values 
  with theoretical calculations from the TAMU and DABMod models. 
Generally, the TAMU model describes many 
  features of the data well, especially the weak \pt\ dependence of
  \RAA, while DABMod only reproduces the measured \RAA\ for $\pT > 12$~\GeV. 
The failure of the DABMod model at low \pt\ is understood
  to result from incomplete modeling of
  heavy-flavor suppression for $\pt \lesssim m_b$.
The TAMU model predicts a larger suppression in
  the 40--60\% centrality interval and a lower suppression in the 0--10\% 
  centrality interval than what is measured.
Thus, the range of the suppression seen in the data is larger than in 
  the TAMU model.
As stated above, the TAMU model does not implement event-by-event fluctuations in the 
  initial geometry, which are known to affect the dynamical evolution of
  bulk medium~\cite{Alver:2006wh,Miller:2007ri}. 
This may be one of the possible reasons for the smaller dynamical range of \RAA\ 
  predicted by the model.

\begin{figure}[!h]
\centerline{
\includegraphics[width=0.9\columnwidth]{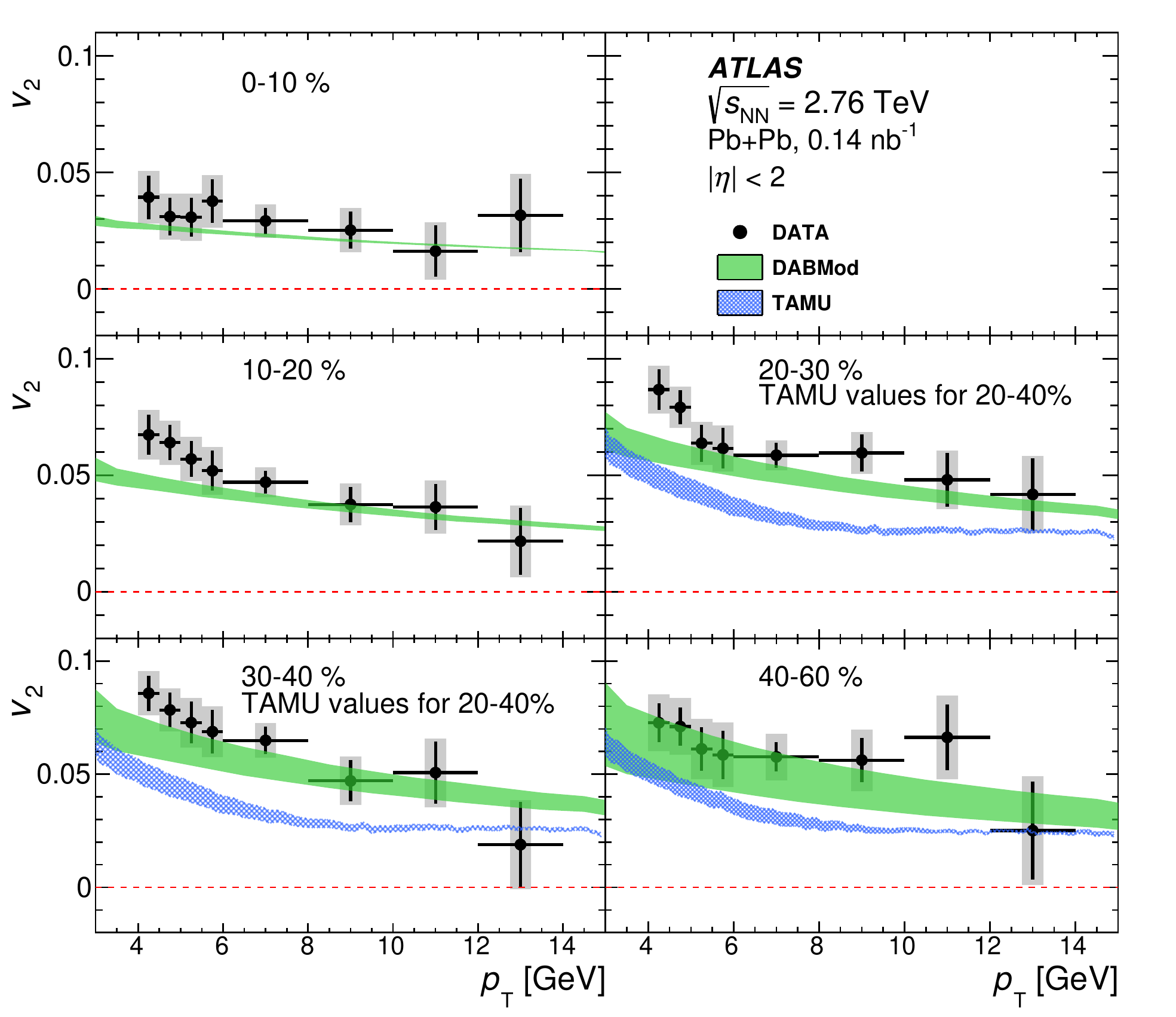}
}
\caption{
Comparison of the \PbPb\ heavy-flavor muon \vtwo\ with calculations 
  from the TAMU and DABMod models.
Each panel represents a different centrality interval.
For the 20--30\% and 30--40\% centrality intervals, the plotted TAMU values 
  correspond to the 20--40\% centrality interval.
For the data, the error bars and shaded bands represent statistical and 
  total uncertainties, respectively. 
For the model calculations, the bands represent theoretical systematic uncertainties.
} 
\label{fig:v2_pt_theory_compare}
\end{figure}

\begin{figure}[!h]
\centerline{
\includegraphics[width=0.9\columnwidth]{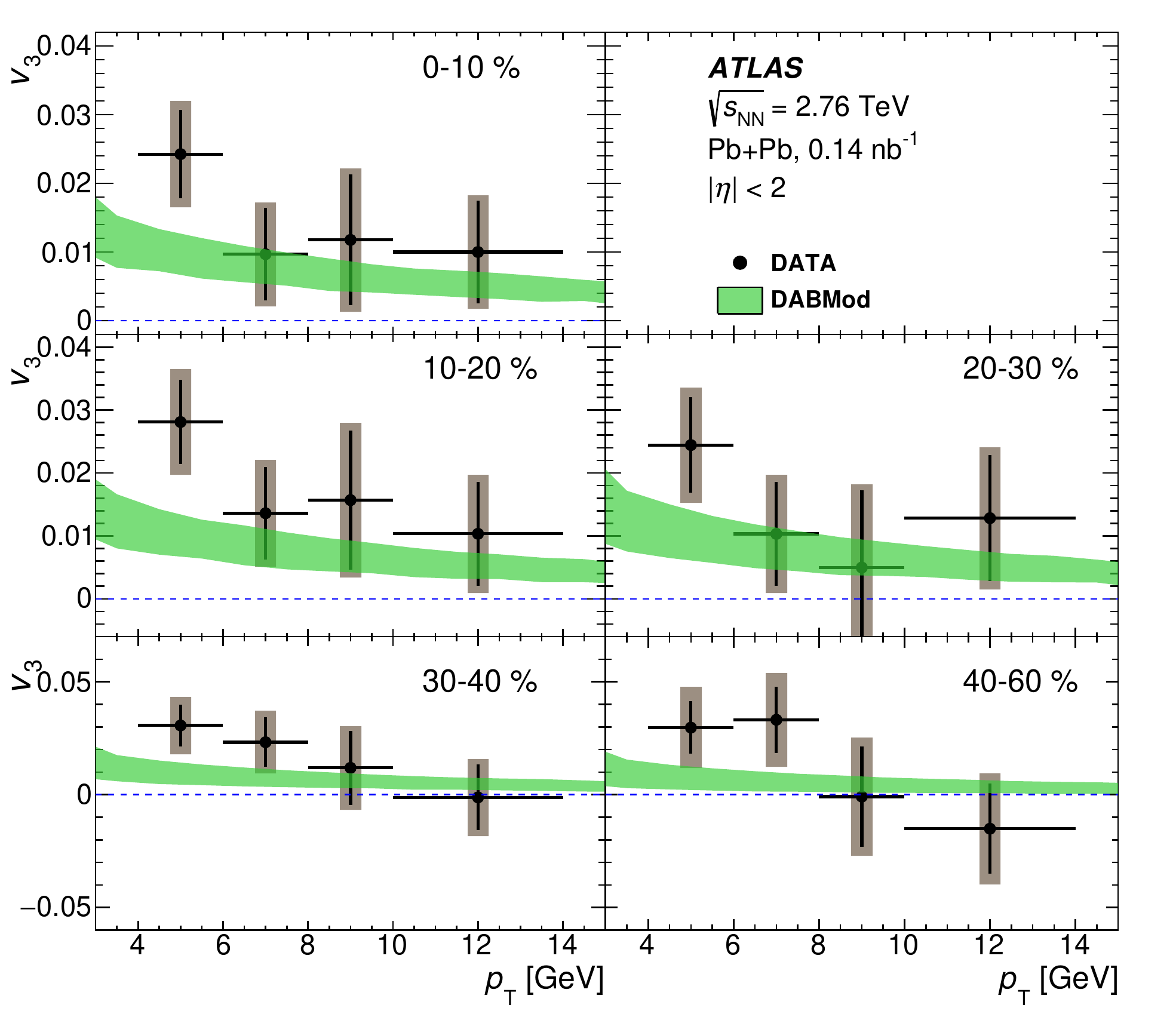}
}
\caption{
Comparison of the \PbPb\ heavy-flavor muon \vthree\ with calculations 
  from the DABMod model.
Each panel represents a different centrality interval.
For the data, the error bars and shaded bands represent statistical and 
  total uncertainties, respectively. 
For the model calculations, the bands represent theoretical systematic uncertainties.
} 
\label{fig:v3_pt_theory_compare}
\end{figure}

Figure~\ref{fig:v2_pt_theory_compare} compares the measured 
  heavy-flavor \vtwo values with calculations from the TAMU 
  and DABMod models.
The DABMod \vtwo\ values are systematically larger than the TAMU values 
  and closer to the measured \vtwo.
Unlike TAMU, the DABMod calculations 
  include event-by-event fluctuations, which are known to increase 
  \vn~\cite{Alver:2006wh,Miller:2007ri}.
This could be a possible reason for the systematically larger \vtwo\ 
  values obtained in the DABMod model.
The DABMod calculations are consistent with the measured values for 
  $\pt > 6$~\GeV\ for all centralities.
However, for $4 < \pt < 6$~\GeV\ and for the 10--40\% centrality range, 
  the calculated values are significantly smaller than the measured \vtwo\ values.
The TAMU \vtwo\ values are significantly smaller than the measured \vtwo\ over the 
  $4 < \pt < 10$~\GeV\ \pt\ range.
Figure~\ref{fig:v3_pt_theory_compare} compares the measured \vthree\ values to
  calculations from the DABMod model.
Features similar to the \vtwo\ comparison are observed;
  the model predictions are smaller than the measured \vthree\ for $4 < \pt < 6$~\GeV\
  but become consistent with the data at higher \pt.
The DABMod calculations are also compared with the \vfour\ measurements. 
However, the large experimental uncertainties do not allow detailed 
  comparisons with the model predictions.

\FloatBarrier
\section{Conclusion}
\label{sec:conclusion}
This \papertype\ presents ATLAS measurements of heavy-flavor muon 
  production in 0.14~\inb\ of $\sqn = 2.76$~\TeV\ \PbPb\ collisions and 570~\inb\ of
  $\sqs = 2.76$~\TeV\ \pp\ collisions at the LHC.
The measurements are performed over the transverse momentum range of 
  $4 < \pT < 14$~\GeV. 
Backgrounds arising from in-flight pion and kaon decays, hadronic showers, 
  and mis-reconstructed muons are statistically removed using a 
  template-fitting procedure based on the relative difference 
  between the muon track momenta in the muon spectrometer and inner detector, 
  corrected for energy loss in the calorimeter system. 
The heavy-flavor muon differential cross-sections and per-event yields 
  are measured in \pp\ and \PbPb\ collisions, respectively. 
The nuclear modification factor \RAA\ calculated from these quantities shows 
  a centrality-dependent suppression that does not depend on \pT\ within uncertainties. 
In the 0--10\% centrality interval, $\RAA \sim0.35$. 
 In \PbPb\ collisions, measurements of the heavy-flavor muon yields as 
  a function of $\phi - \Psi_n$, the azimuthal angle of the muons relative to 
  the event-plane angles, show a clear sinusoidal modulation of the yield in all 
  centrality intervals. 
The heavy-flavor muon \vn, for $n$ = 2--4, is measured in \PbPb\ collisions as a function of \pT\ for 
  five centrality intervals covering the 0--60\% centrality range. 
Significant \vtwo\ values up to about 0.08 are observed at $\pt = 4$~\GeV.
In the 10--20\%, 20--30\%, and 30--40\% intervals, the \vtwo\ decreases with \pT 
  but is still significant at 10~\GeV. 
At fixed \pT, the \vtwo\ values show a systematic variation with centrality 
  which is typical of elliptic-flow measurements.
For most centrality intervals, \vthree\ also decreases with increasing \pt\ 
  over the 4--8~\GeV\ \pt\ range. 
For $\pt > 8$~\GeV, the statistical uncertainties in the measured \vthree\ values are too large
  to discern any \pt-dependent trends.
At a given \pt\ and centrality, the \vthree\ values are smaller than the \vtwo\
  values by a factor of 2--4.
Further, \vthree\ shows a much weaker centrality dependence than \vtwo.
Conclusions about any \pt- or centrality-dependent trends in the \vfour\ 
  are limited by the statistical precision.

The measured \RAA\ and \vtwo\ are also compared with theoretical predictions from
  the TAMU and DABMod models.
The \RAA\ values from the TAMU model show a weak \pt\ dependence over the 
  4--14~\GeV\ \pt\ range, qualitatively similar to the measured \RAA. 
However, the predicted \RAA\ values are smaller than the measured values in the 40--60\%
  centrality interval, and larger than the measured values in the 0--10\% centrality interval.
On the other hand, the DABMod model predicts a strong \pt\ dependence for \RAA,
  which is not observed in the data.
The \RAA\ value at \pt\ = 4~\GeV\ predicted by DABMod is significantly smaller than the measured values
  but increases with increasing \pt\ and becomes comparable to the measured values at 
  \pt\ = 12~\GeV.
For \vtwo, the TAMU and DABMod qualitatively reproduce the observed \pt\ dependence
  but the DABMod calculations are more consistent with the measured values.
Thus both models fail to simultaneously reproduce  \vtwo\ and \RAA\ over 
  the measured \pt\ range.

The \RAA\ values measured here for $|\eta| < 1$ and \vtwo\ values for $|\eta| < 2$, 
  are compatible with, but are substantially more precise than, similar measurements 
  of heavy-flavor muons at forward rapidity ($2.5 < y < 4$) 
  and heavy-flavor electrons at mid-rapidity ($|y| < 0.6$) from the ALICE Collaboration.
Thus, they should provide improved insight into the propagation of heavy quarks in the
  quark--gluon plasma created in \PbPb\ collisions.

\section*{Acknowledgments}

We thank CERN for the very successful operation of the LHC, as well as the
support staff from our institutions without whom ATLAS could not be
operated efficiently.

We acknowledge the support of ANPCyT, Argentina; YerPhI, Armenia; ARC, Australia; BMWFW and FWF, Austria; ANAS, Azerbaijan; SSTC, Belarus; CNPq and FAPESP, Brazil; NSERC, NRC and CFI, Canada; CERN; CONICYT, Chile; CAS, MOST and NSFC, China; COLCIENCIAS, Colombia; MSMT CR, MPO CR and VSC CR, Czech Republic; DNRF and DNSRC, Denmark; IN2P3-CNRS, CEA-DRF/IRFU, France; SRNSFG, Georgia; BMBF, HGF, and MPG, Germany; GSRT, Greece; RGC, Hong Kong SAR, China; ISF, I-CORE and Benoziyo Center, Israel; INFN, Italy; MEXT and JSPS, Japan; CNRST, Morocco; NWO, Netherlands; RCN, Norway; MNiSW and NCN, Poland; FCT, Portugal; MNE/IFA, Romania; MES of Russia and NRC KI, Russian Federation; JINR; MESTD, Serbia; MSSR, Slovakia; ARRS and MIZ\v{S}, Slovenia; DST/NRF, South Africa; MINECO, Spain; SRC and Wallenberg Foundation, Sweden; SERI, SNSF and Cantons of Bern and Geneva, Switzerland; MOST, Taiwan; TAEK, Turkey; STFC, United Kingdom; DOE and NSF, United States of America. In addition, individual groups and members have received support from BCKDF, the Canada Council, CANARIE, CRC, Compute Canada, FQRNT, and the Ontario Innovation Trust, Canada; EPLANET, ERC, ERDF, FP7, Horizon 2020 and Marie Sk{\l}odowska-Curie Actions, European Union; Investissements d'Avenir Labex and Idex, ANR, R{\'e}gion Auvergne and Fondation Partager le Savoir, France; DFG and AvH Foundation, Germany; Herakleitos, Thales and Aristeia programmes co-financed by EU-ESF and the Greek NSRF; BSF, GIF and Minerva, Israel; BRF, Norway; CERCA Programme Generalitat de Catalunya, Generalitat Valenciana, Spain; the Royal Society and Leverhulme Trust, United Kingdom.

The crucial computing support from all WLCG partners is acknowledged gratefully, in particular from CERN, the ATLAS Tier-1 facilities at TRIUMF (Canada), NDGF (Denmark, Norway, Sweden), CC-IN2P3 (France), KIT/GridKA (Germany), INFN-CNAF (Italy), NL-T1 (Netherlands), PIC (Spain), ASGC (Taiwan), RAL (UK) and BNL (USA), the Tier-2 facilities worldwide and large non-WLCG resource providers. Major contributors of computing resources are listed in Ref.~\cite{ATL-GEN-PUB-2016-002}.

\printbibliography

\clearpage 
 
\begin{flushleft}
{\Large The ATLAS Collaboration}

\bigskip

M.~Aaboud$^\textrm{\scriptsize 34d}$,    
G.~Aad$^\textrm{\scriptsize 99}$,    
B.~Abbott$^\textrm{\scriptsize 124}$,    
O.~Abdinov$^\textrm{\scriptsize 13,*}$,    
B.~Abeloos$^\textrm{\scriptsize 128}$,    
D.K.~Abhayasinghe$^\textrm{\scriptsize 91}$,    
S.H.~Abidi$^\textrm{\scriptsize 165}$,    
O.S.~AbouZeid$^\textrm{\scriptsize 143}$,    
N.L.~Abraham$^\textrm{\scriptsize 153}$,    
H.~Abramowicz$^\textrm{\scriptsize 159}$
% ...
\bigskip
\\

% ...
$^{13}$Institute of Physics, Azerbaijan Academy of Sciences, Baku; Azerbaijan.\\
% ...
$^{34}$$^{(a)}$Facult\'e des Sciences Ain Chock, R\'eseau Universitaire de Physique des Hautes Energies - Universit\'e Hassan II, Casablanca;$^{(b)}$Centre National de l'Energie des Sciences Techniques Nucleaires (CNESTEN), Rabat;$^{(c)}$Facult\'e des Sciences Semlalia, Universit\'e Cadi Ayyad, LPHEA-Marrakech;$^{(d)}$Facult\'e des Sciences, Universit\'e Mohamed Premier and LPTPM, Oujda;$^{(e)}$Facult\'e des sciences, Universit\'e Mohammed V, Rabat; Morocco.\\
% ...
$^{91}$Department of Physics, Royal Holloway University of London, Egham; United Kingdom.\\
% ...
$^{99}$CPPM, Aix-Marseille Universit\'e, CNRS/IN2P3, Marseille; France.\\
% ...
$^{124}$Homer L. Dodge Department of Physics and Astronomy, University of Oklahoma, Norman OK; United States of America.\\
% ...
$^{128}$LAL, Universit\'e Paris-Sud, CNRS/IN2P3, Universit\'e Paris-Saclay, Orsay; France.\\
% ...
$^{143}$Santa Cruz Institute for Particle Physics, University of California Santa Cruz, Santa Cruz CA; United States of America.\\
% ...
$^{153}$Department of Physics and Astronomy, University of Sussex, Brighton; United Kingdom.\\
% ...
$^{159}$Raymond and Beverly Sackler School of Physics and Astronomy, Tel Aviv University, Tel Aviv; Israel.\\
% ...
$^{165}$Department of Physics, University of Toronto, Toronto ON; Canada.\\
% ...
$^{a}$ Also at  Department of Physics, University of Malaya, Kuala Lumpur; Malaysia.\\
$^{b}$ Also at Borough of Manhattan Community College, City University of New York, NY; United States of America.\\
$^{c}$ Also at California State University, East Bay; United States of America.\\
$^{d}$ Also at Centre for High Performance Computing, CSIR Campus, Rosebank, Cape Town; South Africa.\\
$^{e}$ Also at CERN, Geneva; Switzerland.\\
% ...
$^{*}$ Deceased

\end{flushleft}


\end{document}